\documentclass[prd,preprint,tightenlines,floatfix, showpacs,preprintnumbers,nofootinbib,eqsecnum,superscriptaddress]{revtex4}

\usepackage[T1]{fontenc}		

\usepackage{amsmath,amsfonts,amssymb,amstext,mathrsfs}
\usepackage{mathpazo}

\usepackage[dvips]{graphicx}
\usepackage{epsf,float}
\usepackage{revsymb}

\usepackage{dcolumn}
\usepackage{braket}
\usepackage{color,xcolor}
\usepackage{graphicx}
\usepackage{subfigure}
\usepackage{multirow}
\usepackage{tabularx}
\usepackage{pstricks}
\usepackage[section]{placeins}
\usepackage{booktabs}
\usepackage{array}

\usepackage{hyperref}

\usepackage{subfigure}

\newcommand{\Pom}{\mathbb{P}}

\newcommand{\Reg}{\mathbb{R}}

\begin{document}


\title{Triple Regge exchange mechanisms \\ of four-pion continuum production \\ in the $pp \to pp \pi^{+}\pi^{-}\pi^{+}\pi^{-}$ reaction}

\author{Rados{\l}aw Kycia\footnote{Also at The Faculty of Science, Masaryk University, Kotl\'{a}\v{r}sk\'{a} 2, 602 00 Brno, Czechia.}}
 \email{kycia.radoslaw@gmail.com}
\affiliation{Cracow University of Technology, Faculty of Physics, Mathematics and Computer Science, PL-31155, Krak\'ow, Poland}

\author{Piotr Lebiedowicz}
 \email{Piotr.Lebiedowicz@ifj.edu.pl}
\affiliation{Institute of Nuclear Physics Polish Academy of Sciences, PL-31342 Krak\'ow, Poland}

\author{Antoni Szczurek
\footnote{Also at The Faculty of Mathematics and Natural Sciences, University of Rzeszow, ul.
Pigonia 1, 35-310 Rzeszow, Poland.}}
\email{Antoni.Szczurek@ifj.edu.pl}
\affiliation{Institute of Nuclear Physics Polish Academy of Sciences, PL-31342 Krak\'ow, Poland}

\author{Jacek Turnau}


\begin{abstract}
\noindent
We consider exclusive multi-peripheral production of four charged pions
in proton-proton collisions at high energies with simultaneous exchange
of three pomerons/reggeons. The amplitude(s) for the genuine $2 \to 6$ process
are written in the Regge approach. The calculation is performed
with the help of the GenEx Monte Carlo code.
Some corrections at low invariant masses in the two-body subsystems
are necessary for application of the Regge formalism.
We estimate the corresponding cross section
and present differential distributions in rapidity, transverse momenta
and two- and four-pion invariant masses. The cross section and 
the distributions depend on the value of the cut-off parameter
of a form factor correcting amplitudes for off-shellness of $t$-channel pions.
Rather large cross section is found for the whole phase space
($\sigma \sim$ 1-5 $\mu$b, including absorption corrections). Relatively large four-pion invariant masses
are populated in the considered diffractive mechanism compared
to other mechanisms discussed so far in the context of four-pion production.
We investigate whether the triple Regge exchange processes could be identified with
the existing LHC detectors. We consider the case of ATLAS and ALICE cuts. The ATLAS (or CMS) has better chances
to identify the process in the region of large invariant masses
$M_{4 \pi} > 10$ GeV. In the case of the ALICE experiment 
the considered mechanism competes with other mechanisms
(production of $\sigma \sigma$, $\rho \rho$ pairs or single resonances) and cannot be unambiguously identified.
\end{abstract}


\maketitle

\section{Introduction}

In the present paper, we study the exclusive $2 \to 6$ process: 
\begin{equation}
p p \to p p \pi^+ \pi^- \pi^+ \pi^- \,.
\label{our_reaction}
\end{equation}
In general, the number of possible mechanisms is rather large.
Here we shall focus on the triple-Regge exchange processes.
According to our knowledge such processes were not discussed
quantitatively in the literature and estimation of their importance
becomes timely in the light of studies being performed
by the STAR, ATLAS, CMS and ALICE collaborations.
At the LHC the energy is so high that there is enough rapidity span
for such processes to occur, at least from the theoretical point of view.

In this study we present an extension of the Regge-inspired 
Lebiedowicz-Szczurek approach used for the reactions: 
$pp \to pp \pi^+ \pi^-$ \cite{LS_2pi}, \cite{LSSTC_2pi},
$pp \to nn \pi^+ \pi^+$ \cite{LS_nnpipi} and $pp \to pp K^{+} K^{-}$ \cite{LS_2K}.
The number of diagrams for the six-body reactions is bigger than for the four-body reaction and we have 
to carefully write the corresponding amplitudes using, however, simplified Regge rules for $\pi p$ and $\pi \pi$ interactions.

We shall try to use the same model parameters as for
the $p p \to p p \pi^+ \pi^-$ whenever possible.
This should allow for an approximate estimation of the cross section
and some differential distributions.
The calculation presented here is performed
with the help of the GenEx Monte Carlo event generator \cite{GenEx}.

We wish to concentrate on the four charged pion continuum production mechanism
which, in addition, is a background for studies of central exclusive production of resonances
discussed recently in \cite{LNS2016_2pi}.
The production of glueball states is expected to be enhanced
in gluon rich pomeron-pomeron interactions.
Identification of diffractively produced glueball states 
is still an experimental challenge at the LHC.
For experimental point of view at lower energies see e.g. \cite{ABCDHW_4pi}.
This requires calculation/estimation of the four-pion background 
from different sources, see e.g. \cite{LNS2016_4pi}.

\section{Amplitude for the four-pion continuum production}

\begin{figure}[htp] 
\includegraphics[width=5.5cm]{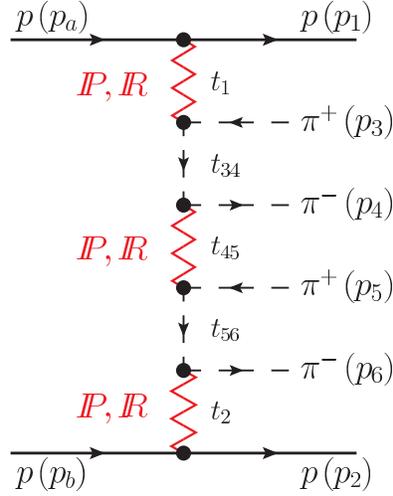}     
\caption{A diagram for exclusive diffractive production of $\pi^+ \pi^-\pi^+ \pi^-$  continuum 
with three pomeron/reggeon exchanges
in proton-proton collisions. Here $\Reg$ denotes the $f_{2 \Reg}$ or $\rho_{\Reg}$ reggeon exchanges. 
The four-momentum transfers squared are shown explicitly.}
\label{fig:diagrams_4pi_Born}
\end{figure}

The general situation for the $pp \to pp \pi^{+}\pi^{-}\pi^{+}\pi^{-}$ process
is sketched in Fig.~\ref{fig:diagrams_4pi_Born}.
The full amplitude, including different permutations of outgoing pion pairs,
can be written as
\footnote{
Here we introduce a shorthand notation that corresponds to the four-momenta of outgoing pions,
e.g. $\{ 3456 \}$ means that
the index 3 is for $\pi^{+}(p_{3})$, index 4 is for $\pi^{-}(p_{4})$,
index 5 is for another $\pi^{+}(p_{5})$,
and index 6 is for another $\pi^{-}(p_{6})$,
see the diagram in Fig.~\ref{fig:diagrams_4pi_Born}.}
\begin{equation}
\begin{split}
{\cal M} &= \frac{1}{2}\left({\cal M}_{\{3456\}} + {\cal M}_{\{5436\}} + {\cal M}_{\{3654\}} 
+ {\cal M}_{\{5634\}}
\right)\\
         &+ \frac{1}{2}\left({\cal M}_{\{4356\}} + {\cal M}_{\{4536\}} + {\cal M}_{\{6354\}} 
+ {\cal M}_{\{6534\}}
\right)\\
         &+ \frac{1}{2}\left({\cal M}_{\{3465\}} + {\cal M}_{\{5463\}} + {\cal M}_{\{3645\}} 
+ {\cal M}_{\{5643\}}
\right)\\
         &+ \frac{1}{2}\left({\cal M}_{\{4365\}} + {\cal M}_{\{4563\}} + {\cal M}_{\{6345\}} 
+ {\cal M}_{\{6543\}}
\right)
          \,,
\label{sum_of_diagrams} 
\end{split}
\end{equation}
where the factor $\frac{1}{2}$ is the symmetry factor for two identical pions\footnote{The symmetry factor is artificially written here instead of the factor $\frac{1}{2!2!}$ in the cross section formula.}.

In formulae below the subsystem energies squared is
\begin{equation}
s_{ij} = M_{ij}^{2} = (p_i + p_j)^2 \,,
\label{s_ij}
\end{equation}
where $p_i$ and $p_j$ are respective four-vectors, and
the formulae of four-momentum transfers squared are
\footnote{Here $p_1, p_2, p_3, p_4, p_5, p_6$ should be treated as outgoing.}
\begin{equation}
\begin{split}
&t_{1} = (p_a-p_1)^2 \,, \\
&t_{2} = (p_b-p_2)^2 \,, \\
&t_{34} = (p_a-p_1-p_3)^2 \,, \\
&t_{43} = (p_a-p_1-p_4)^2 \,, \\
&t_{45} = t_{35} = t_{46} = t_{36 }= (p_a-p_1-p_3-p_4)^2 = (p_b-p_2-p_6-p_5)^2 \;, \\
&t_{56} = (p_b-p_2-p_6)^2 \,, \\
&t_{65} = (p_b-p_2-p_5)^2 \,.
\label{t_definitions}
\end{split}
\end{equation}

We write the amplitude for each group in Eq.~(\ref{sum_of_diagrams}):
\begin{eqnarray}
{\cal M}_{\{3456\}} &=& 
A_{\pi p}(s_{13},t_{1}) \, \frac{F_{\pi}(t_{34})}{t_{34} - m_{\pi}^2} \,
A_{\pi \pi}(s_{45},t_{45}) \, \frac{F_{\pi}(t_{56})}{t_{56} - m_{\pi}^2} 
\, A_{\pi p}(s_{26},t_{2})\,,
\label{amplitude_a}\\
{\cal M}_{\{4356\}} &=& 
A_{\pi p}(s_{14},t_{1}) \, \frac{F_{\pi}(t_{43})}{t_{43} - m_{\pi}^2} \,
A_{\pi \pi}(s_{35},t_{35}) \, \frac{F_{\pi}(t_{56})}{t_{56} - m_{\pi}^2} 
\, A_{\pi p}(s_{26},t_{2})\,,
\label{amplitude_b}\\
{\cal M}_{\{3465\}} &=& 
A_{\pi p}(s_{13},t_{1}) \, \frac{F_{\pi}(t_{34})}{t_{34} - m_{\pi}^2} \,
A_{\pi \pi}(s_{46},t_{46}) \, \frac{F_{\pi}(t_{65})}{t_{65} - m_{\pi}^2} 
\, A_{\pi p}(s_{25},t_{2})\,,
\label{amplitude_c}\\
{\cal M}_{\{4365\}} &=& 
A_{\pi p}(s_{14},t_{1}) \, \frac{F_{\pi}(t_{43})}{t_{43} - m_{\pi}^2} \,
A_{\pi \pi}(s_{36},t_{36}) \, \frac{F_{\pi}(t_{65})}{t_{65} - m_{\pi}^2} 
\, A_{\pi p}(s_{25},t_{2})\,.
\label{amplitude_d}
\end{eqnarray}
The subprocess amplitudes with the Regge exchanges are given as
\begin{eqnarray}
&& A_{\pi p}(s,t) = 
\sum_{j = \Pom, f_{2 \Reg}}\eta_{j} \,s\,C_{\pi p}^{j} \left( \frac{s}{s_0} \right)^{\alpha_{j}(t)-1}
F_{\pi p}^{j}(t)\,,
\label{amplitude_for_ppi_subproces}\\
&&A_{\pi \pi}(s,t) =
\sum_{j = \Pom, f_{2 \Reg}} \eta_{j} \,s\,C_{\pi \pi}^{j} \left( \frac{s}{s_0} \right)^{\alpha_{j}(t)-1}
F_{\pi \pi}^{j}(t)\,,
\label{amplitude_for_pipi_subproces}
\end{eqnarray}
where the signature factors at $t=0$ are $\eta_{\Pom} = i$ and 
$\eta_{f_{2 \Reg}} = i - 0.86$ \cite{LS_2pi}.
The interaction strength parameters are assumed to fulfil 
the Regge factorization relation:
\begin{equation} 
C_{pp}^{j} C_{\pi \pi}^{j} = C_{\pi p}^{j} C_{\pi p}^{j}\,,
\label{factorization}
\end{equation}
where $j = \Pom, f_{2 \Reg}$.
In our calculations we use the following numerical parameters
\begin{eqnarray}
&&C_{p p}^{\Pom}   = 21.70\;{\rm mb} \,, \quad
C_{\pi p}^{\Pom}   = 13.63\;{\rm mb} \,, \quad
C_{\pi \pi}^{\Pom} = 8.56\;{\rm mb} \,,
\label{pom_coupl_param}\\
&&C_{p p}^{f_{2\Reg}}   = 75.4875\;{\rm mb} \,, \quad
C_{\pi p}^{f_{2\Reg}}   = 31.79\;{\rm mb} \,, \quad
C_{\pi \pi}^{f_{2\Reg}} = 13.39\;{\rm mb} \,.
\label{f2reg_coupl_param}
\end{eqnarray}

We parametrize the $t$-dependences of subprocess amplitudes
in the exponential form:
\begin{eqnarray}
F_{\pi p}^{j}(t) &=&   \exp \left( \frac{B_{\pi p}^{j}}{2} t   \right) \,, \\
F_{\pi \pi}^{j}(t) &=& \exp \left( \frac{B_{\pi \pi}^{j}}{2} t \right) \,,
\label{t_dependences_of_Regge_amplitudes}
\end{eqnarray}
where the slope parameters are taken as 
$B_{\pi p}^{\Pom} = 5.5$~GeV$^{-2}$, $B_{\pi \pi}^{\Pom} = 4$~GeV$^{-2}$, 
$B_{\pi p}^{f_{2 \Reg}} = 4$~GeV$^{-2}$, $B_{\pi \pi}^{f_{2 \Reg}} = 4$~GeV$^{-2}$ (see \cite{LS_2pi}).

The Regge trajectories $\alpha_{j}(t)$ are assumed to be of the standard linear form \cite{DDLN}:
\begin{eqnarray}
&&\alpha_{j}(t) = \alpha_{j}(0) + \alpha'_{j}t\,,
\label{trajectory}\\
&&\alpha_{\Pom}(0) = 1.0808\,, \quad 
\alpha'_{\Pom} = 0.25\;{\rm GeV^{-2}} \,,
\label{pomeron_trajectory}\\
&&\alpha_{f_{2 \Reg}}(0) = 0.5475\,, \quad 
\alpha'_{f_{2 \Reg}} = 0.9\;{\rm GeV^{-2}} \,.
\label{reggeon_trajectory}
\end{eqnarray}

The off-shellness of $t$-channel pions in the diagrams is included via 
multiplication of corresponding amplitudes by the extra form factor:
\begin{eqnarray}
F_{\pi}(t) = \exp\left( \frac{t-m_{\pi}^{2}}{\Lambda_{off,E}^2} \right)
=  \exp\left( \frac{t-m_{\pi}^{2}}{2\Lambda_{off,E}^2} \right) 
   \exp\left( \frac{t-m_{\pi}^{2}}{2\Lambda_{off,E}^2} \right)  \,.
\label{off-shell_formfactor}
\end{eqnarray}
In fact the off-shell effects are related to vertices and they
always go in pairs for our process.
The form factor is normalized to unity when meson is on-mass-shell 
$F_{\pi}(m_{\pi}^{2}) = 1$. 
The parameter of the off-shell form factor(s) is in principle 
a free parameter. 
In the present paper we shall use $\Lambda_{off,E} = 1$ GeV (lower limit) and use $\Lambda_{off,E} = 1.5$ GeV (upper limit).
These values correspond to $\tilde{\Lambda}_{off,E} = 1.41$ GeV and $\tilde{\Lambda}_{off,E} = 2.12$ GeV in the convention used in \cite{LS_2pi}.

The amplitudes (\ref{amplitude_for_ppi_subproces}) and (\ref{amplitude_for_pipi_subproces}) 
have to be corrected (cut off) for low
$\sqrt{s_{ij}} = W_{ij}$ as the Regge theory is valid only above a lower subenergy limit. 
In our analysis here mainly a smooth cut-off function will be used, 
as in \cite{LS_2pi}, e.g.,
\begin{equation}
 f_{cont}(W_{ij})=\frac{\exp\left((W_{ij}-W_{0})/a\right)}{1+\exp\left((W_{ij}-W_{0})/a\right)},
\label{smoothCut}
\end{equation}
with $a=0.2$~GeV and $W_{0} = 2$~GeV, which cuts off $s_{ij} \lesssim 4$~GeV$^{2}$ smoothly.

Another cut-off function which we use is the Heaviside theta function:
\begin{equation}
 f_{discont}(W_{ij})=\theta( W_{ij}-W_{cut}),
 \label{thetaCut}
\end{equation}
where $W_{cut}$ is a parameter to be adjusted to future precise data. 
We will show that both functions, see (\ref{smoothCut}) and (\ref{smoothCut}), give similar results for the integrated cross section, however, 
somewhat different distributions in some special variables.

\section{Feasibility study for the measurement of the triple Regge exchange process at current LHC experiments}

In this section we show some  predictions for the considered process. 
We will select $\sqrt{s}= 7$~TeV collision energy as a representative example. 
The collision energy dependence of the cross section is rather weak (see section \ref{Subsection:13TeV}).

Presentation of our results is divided into three parts:\\
(A) calculation for the full six-body phase space, \\ 
(B) calculation relevant for the ATLAS main tracker, \\ 
(C) calculation relevant for the ALICE main tracker. 

In a separate section we discuss some general specific aspects of the discussed here mechanism.

Some technical details related to the Monte Carlo integration are described in Appendix \ref{A}.

\subsection{Results for the full phase space}
\label{subsection:Resultsforthefullphasespace}

In this subsection we present some results for cross section calculation which we call "full phase space" meaning that only minimal cuts are
imposed for purely technical reasons, namely
\begin{equation}
  p_{t,p} < 2 \; {\rm GeV}\,, \quad
  |y_{4\pi}| < 6\,, \quad  M_{4\pi}<30{\rm ~GeV}.
 \label{FullPhaseSpace}
\end{equation}

These cuts can be easily placed in experimental analyses and do not change the shape of the resulting distributions. The condition $ M_{4\pi}<30{\rm ~GeV}$
cuts off less than a few percent of the cross section.

\begin{table}[!htb]
  \caption{Full phase space cross sections in $\mu$b.
           No absorption effects are included here.}
\centering
  \begin{tabular}{| l | c | c | }
    \hline
    & $\Lambda_{off,E}$ [GeV] & $\sigma$ [$\mu$b]  \\ \hline   
    Symmetrization    & 1.0 & \;\,7.21 \\ \hline
    No symmetrization & 1.0 & \;\,0.82 \\ \hline
    Symmetrization    & 1.5 & 42.86 \\ \hline
    No symmetrization & 1.5 & \;\, 4.30 \\ \hline
  \end{tabular}
  \label{tab:CrossSectionFullPhaseSpace}
\end{table}
In Tab.~\ref{tab:CrossSectionFullPhaseSpace} we present 
numerical results for the cross section integrated over so-defined full phase space.
In the table, 'No symmetrization' means that we take only one 
arbitrarily chosen term for the matrix element 
and omit the symmetrization factor,
that is, ${\cal M} = {\cal M}_{\{3456\}}$ in Eq.~(\ref{sum_of_diagrams}).

\begin{figure}[htp]
  \centering
  \includegraphics[width=0.35\textwidth, angle = -90]{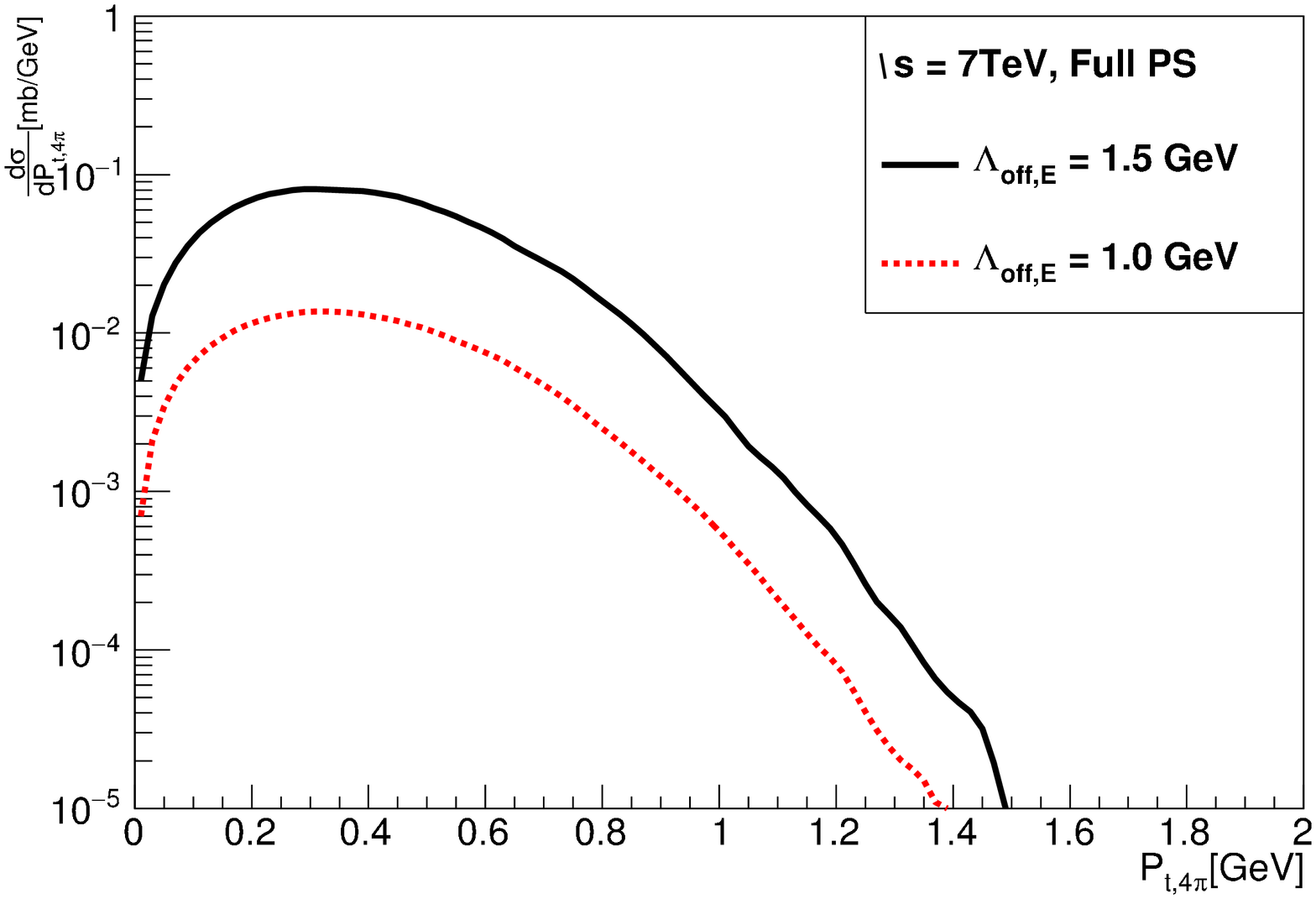}
  \includegraphics[width=0.35\textwidth, angle = -90]{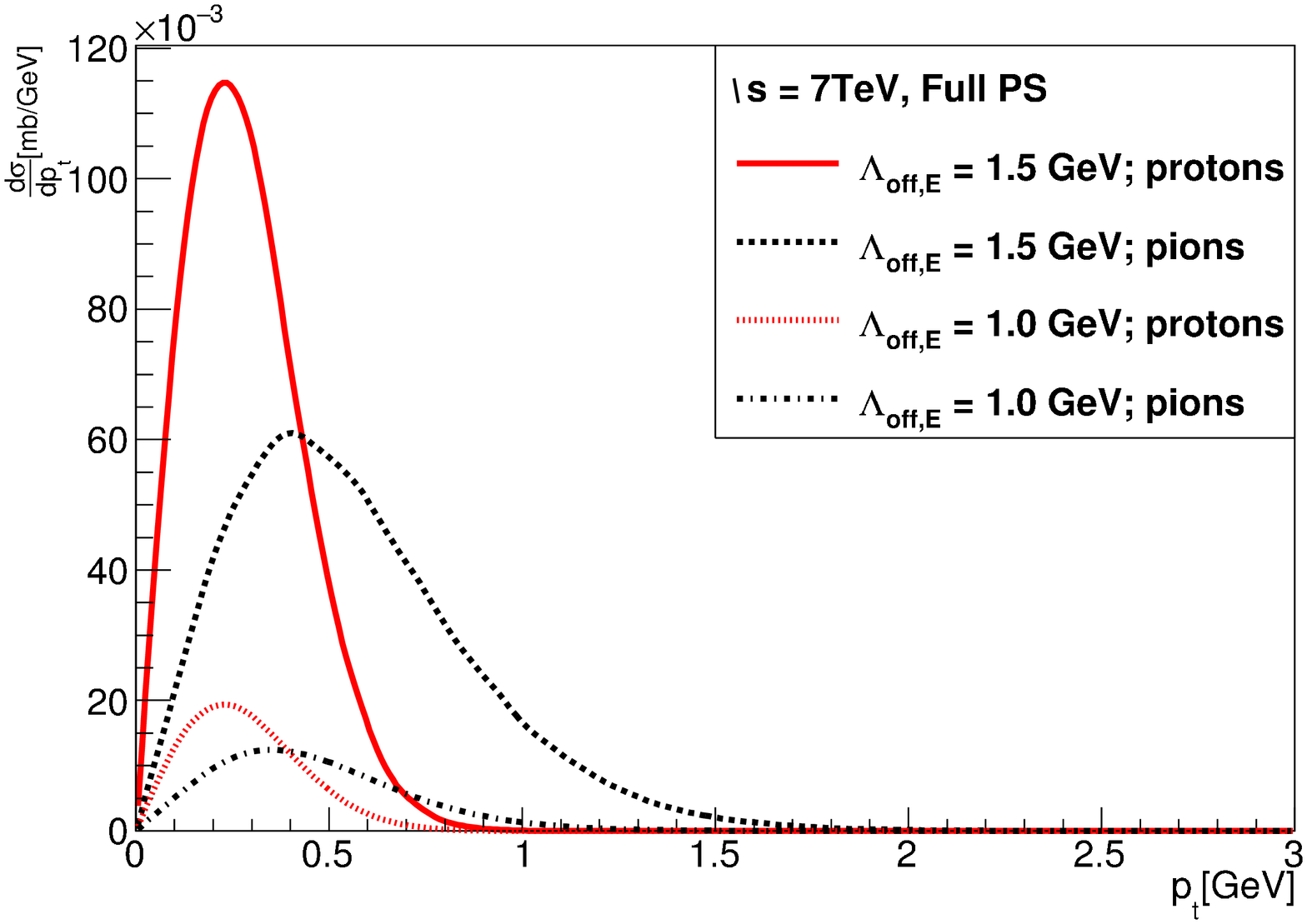}
  \caption{Distributions in transverse momentum of the four-pion system(left panel) and for the transverse momenta of individual particles (protons and pions) for two different values of $\Lambda_{off,E}$ = 1, 1.5 GeV.}  
\label{PH_Pt}
\end{figure}

\begin{figure}[htp]
  \centering
  \includegraphics[width=0.5\textwidth, angle = -90]{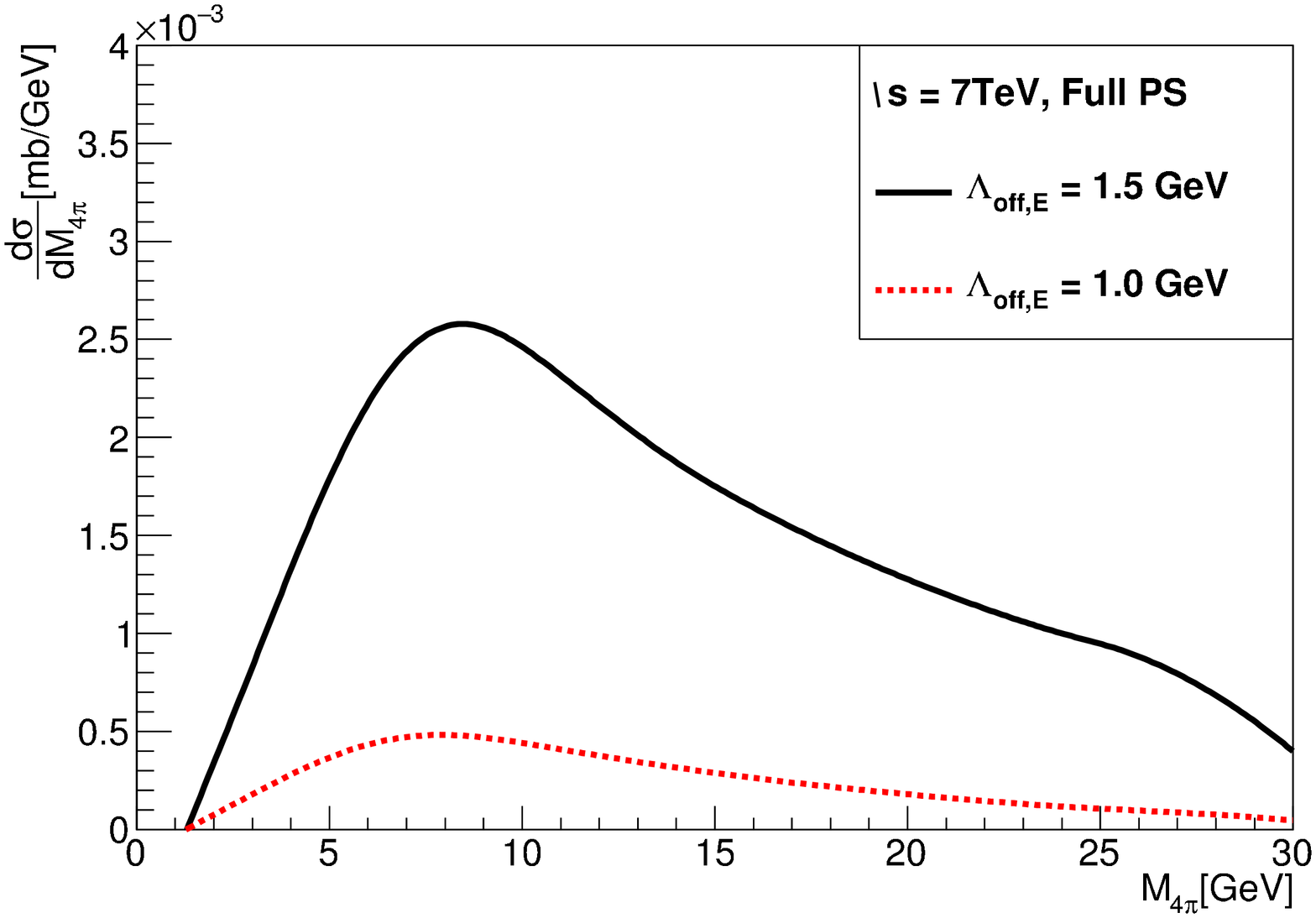}
  \caption{Four-pion invariant mass distribution for the full phase
    space for two different values of $\Lambda_{off,E}$ = 1, 1.5 GeV.
}
  \label{PH_CM}
\end{figure}

\begin{figure}[htp]
  \centering
  \includegraphics[width=0.5\textwidth, angle = -90]{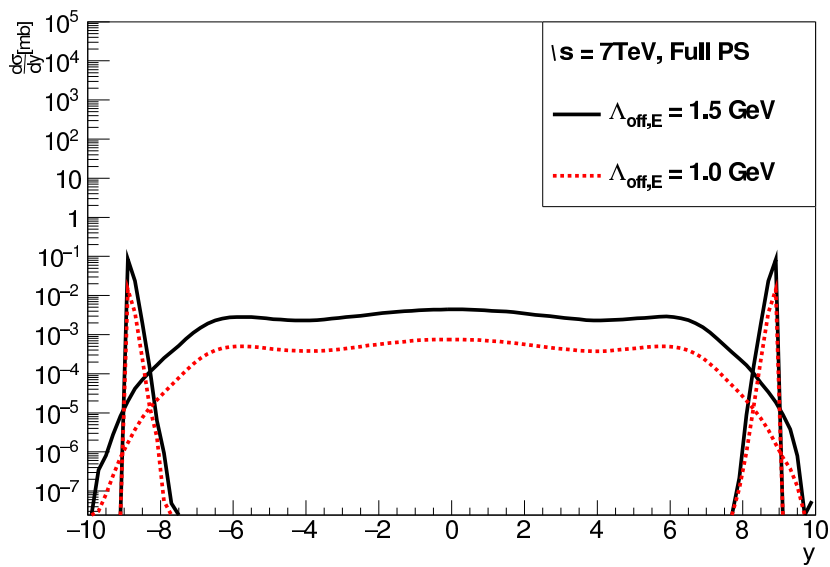}
  \caption{Rapidity distribution of protons (external peaks) and
    pions (internal bumps)  for the full phase space
    for two different values of $\Lambda_{off,E}$ = 1, 1.5 GeV.
}  
\label{PH_Y}
\end{figure}

\begin{figure}[htp]
  \centering
  \includegraphics[width=0.45\textwidth, angle = -90]{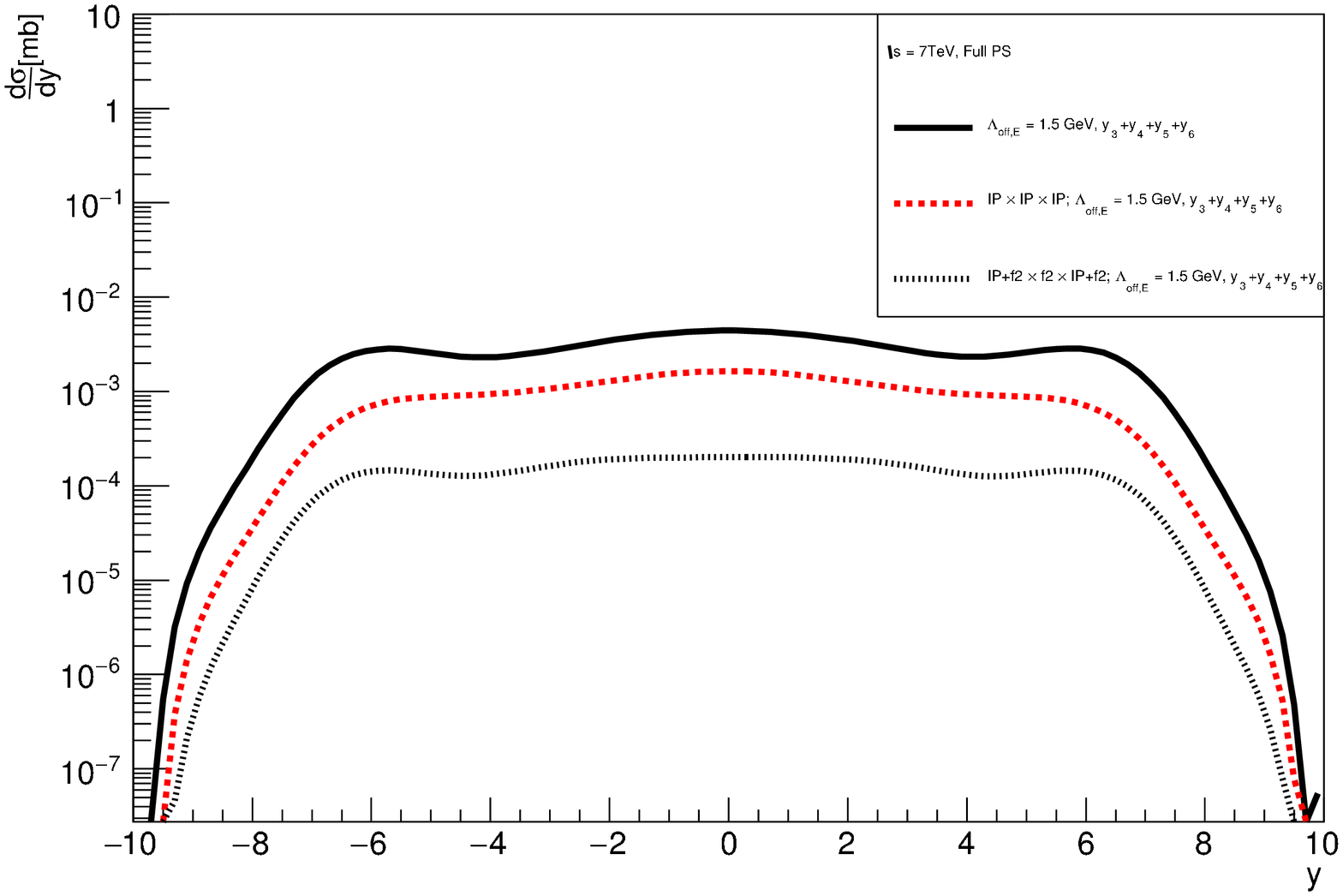}
  \includegraphics[width=0.45\textwidth, angle = -90]{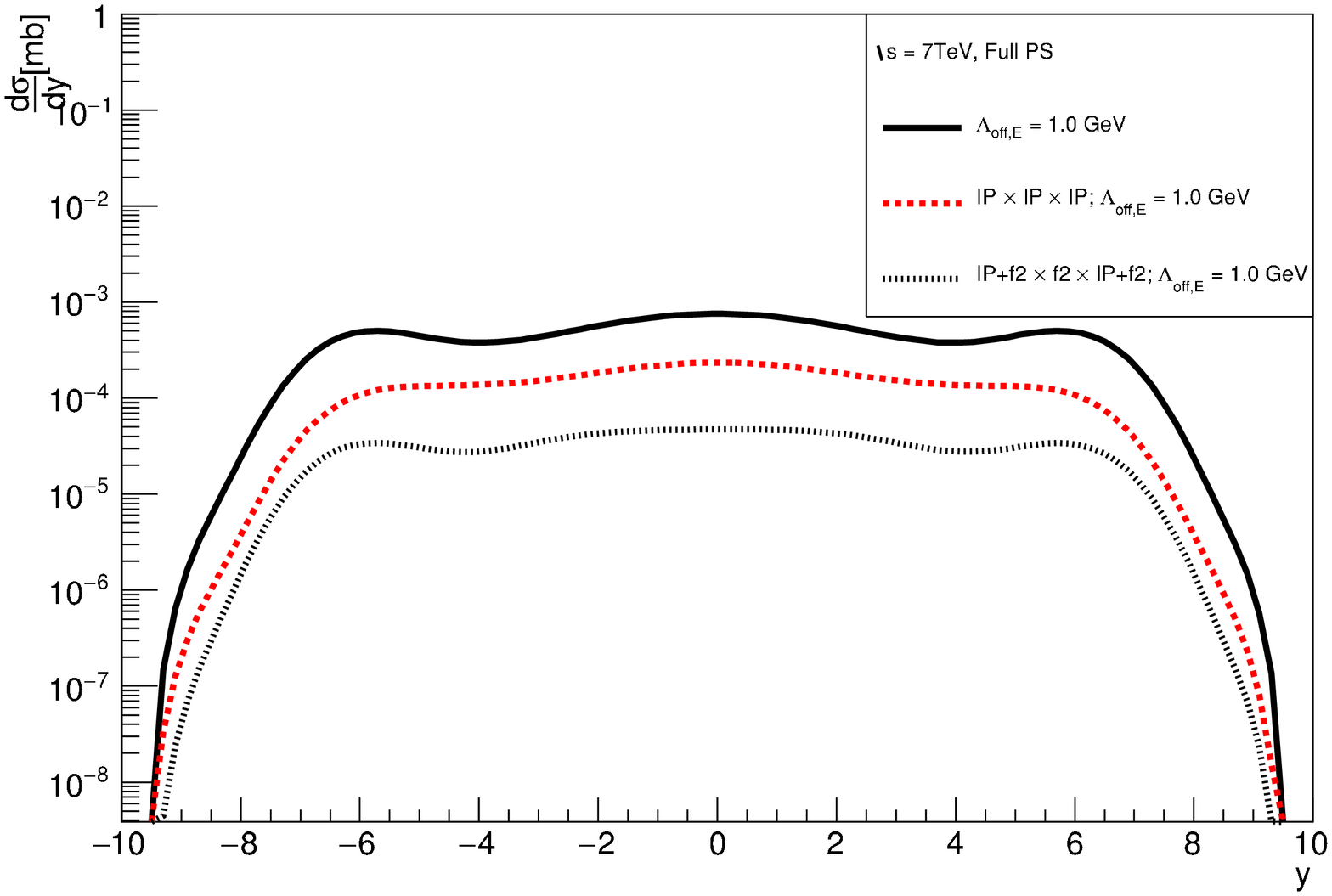}
\caption{Rapidity distribution of pions for 
$(\Pom + f_{2 \Reg}) \times (\Pom + f_{2 \Reg}) \times (\Pom + f_{2 \Reg})$ (upper curve),
$\Pom \times \Pom \times \Pom$ (middle curve) and 
$(\Pom + f_{2 \Reg}) \times f_{2 \Reg}  \times (\Pom + f_{2 \Reg})$ (lower curve)
exchanges for two different values of $\Lambda_{off,E}$ = 1, 1.5 GeV.}
\label{PH_Y_pions}
\end{figure}

\begin{figure}[htp]
  \centering
  \includegraphics[width=0.45\textwidth, angle = -90]{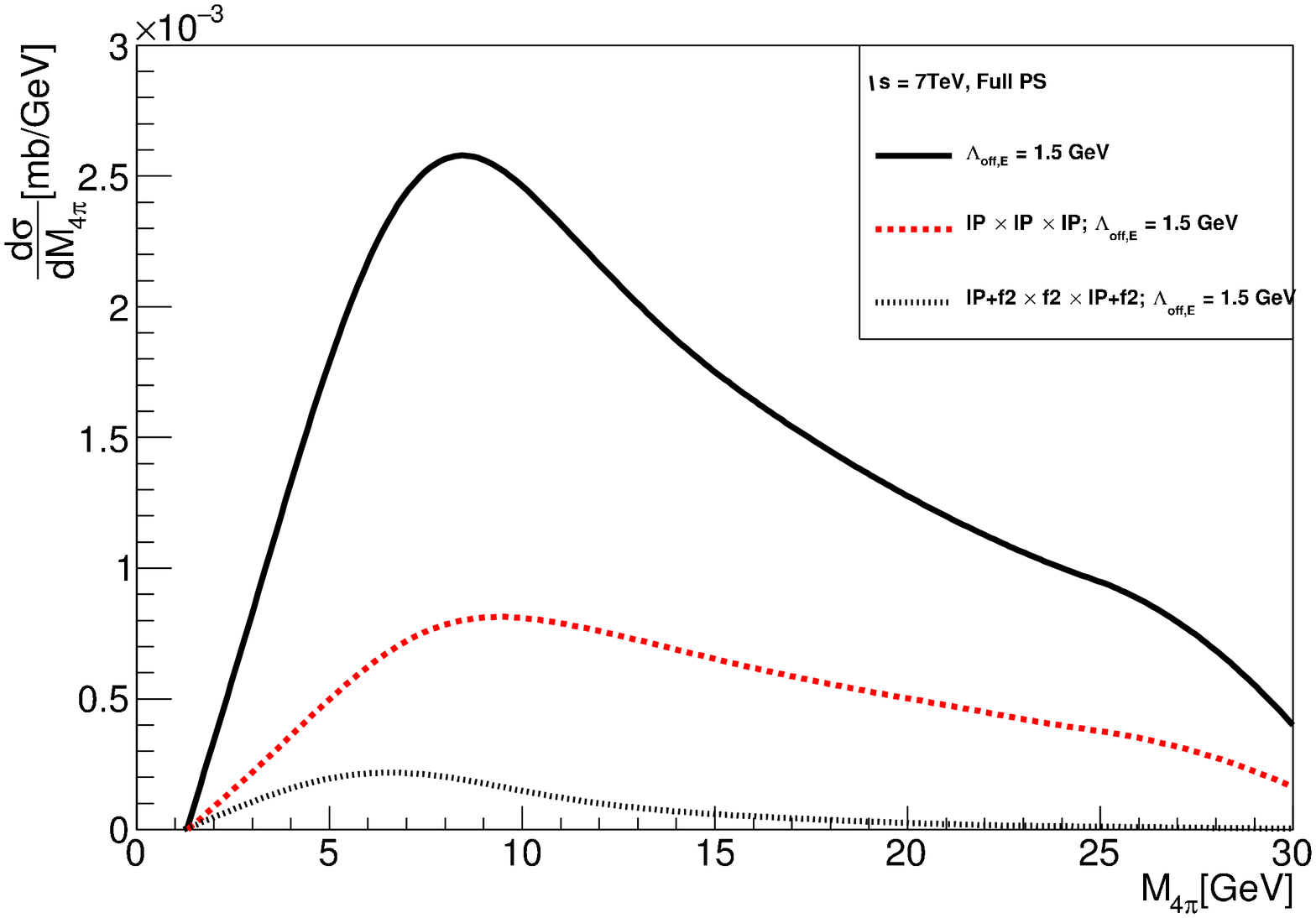}
  \includegraphics[width=0.45\textwidth, angle = -90]{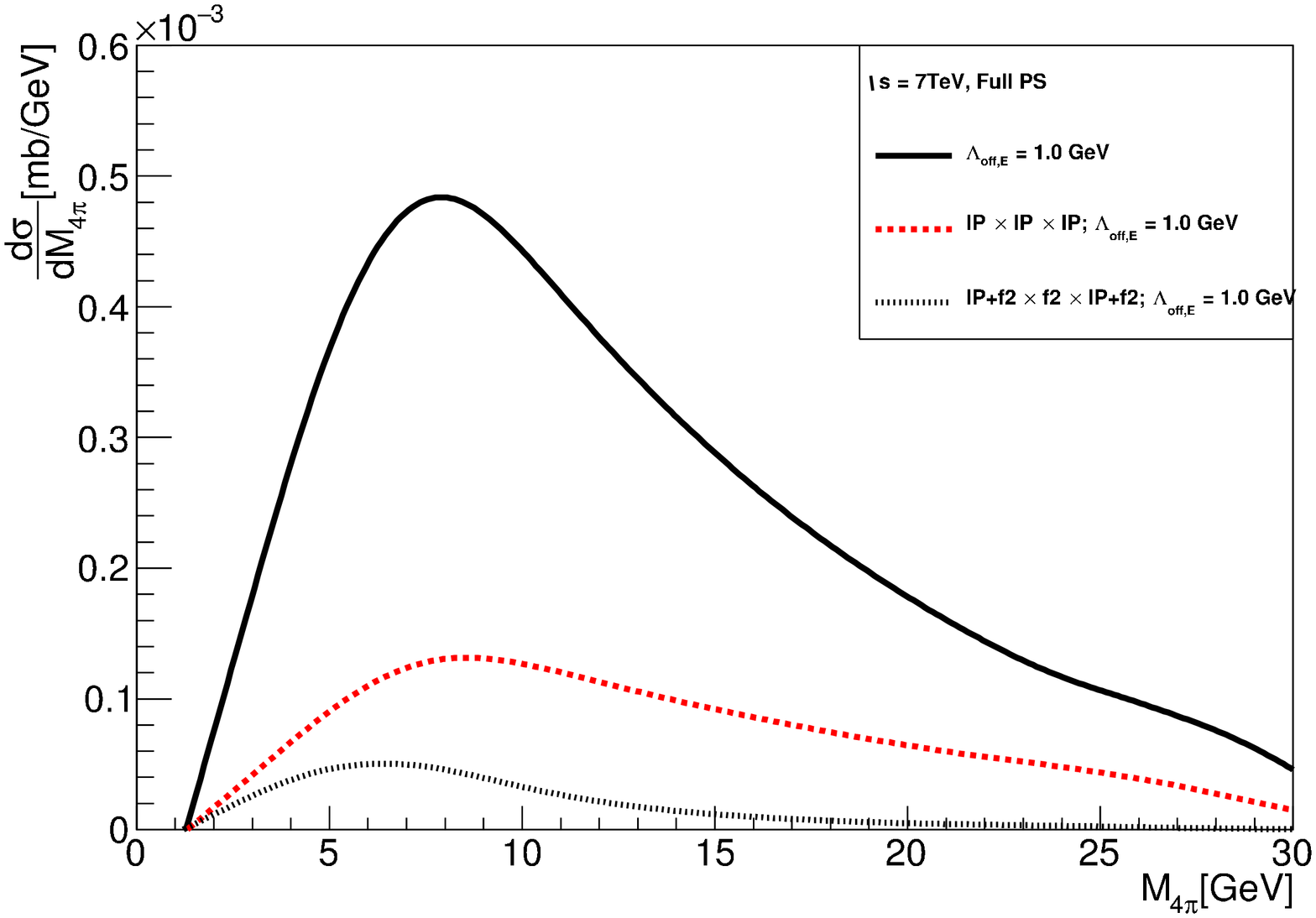}
\caption{$4\pi$ invariant mass distribution for 
$(\Pom + f_{2 \Reg}) \times (\Pom + f_{2 \Reg}) \times (\Pom + f_{2 \Reg})$ (upper curve),
$\Pom \times \Pom \times \Pom$ (middle curve) and 
$(\Pom + f_{2 \Reg}) \times f_{2 \Reg}  \times (\Pom + f_{2 \Reg})$ (lower curve)
exchanges for two different values of $\Lambda_{off,E}$ = 1, 1.5 GeV.}
\label{PH_CM_pionsPomeron}
\end{figure}

\begin{figure}[htp]
  \centering
  \includegraphics[width=0.8\textwidth, angle = -90]{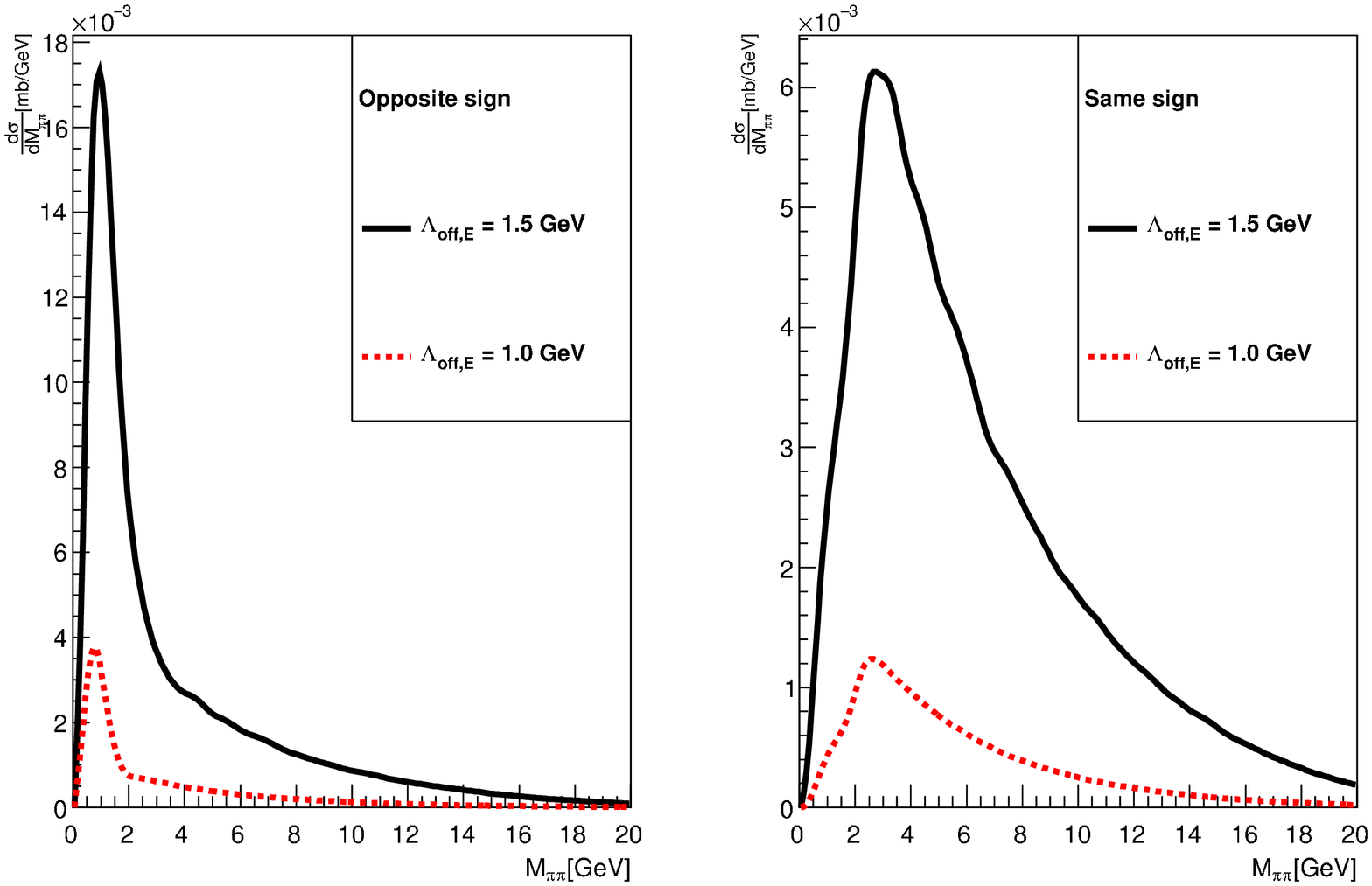}
  \caption{Dipion invariant mass distributions for the opposite-sign
    (left panel) and for the same-sign (right panel) pions
for two different values of $\Lambda_{off,E}$ = 1 GeV 
(lower curve) and $\Lambda_{off,E}$ = 1.5 GeV (upper curve).}  
\label{PH_CMij}
\end{figure}

In Figs. \ref{PH_Pt}, \ref{PH_CM}, \ref{PH_Y} and \ref{PH_CMij} the 
general features of the investigated process are presented while Figs.~\ref{PH_Y_pions} and \ref{PH_CM_pionsPomeron} illustrate relative contributions of the pomeron and subleading reggeon trajectories to the calculated cross sections.

The distribution in four-pion invariant mass, see Fig. \ref{PH_CM}, extends in relatively broad range, compared e.g. to dipion invariant mass distribution for the $p p \to p p \pi^+ \pi^-$ reaction. The four momentum transfer from both protons to the $4\pi$ system is restricted by peripherality of the process what results in relatively narrow distribution of (\ref{pz}) shown in Fig. \ref{pz_distribution}.

At $\sqrt{s} = 7$~TeV the outgoing protons are produced at
$y \approx \pm 9$ (see Fig.~\ref{PH_Y}). The pions 
are produced between protons.
The pion rapidity distribution illustrates well the role of subleading $f_{2 \Reg}$ reggeons.
In Fig.~\ref{PH_Y_pions} we compare distributions for
$(\Pom+f_{2 \Reg}) \times (\Pom+f_{2 \Reg}) \times (\Pom+f_{2 \Reg})$ and $\Pom \times \Pom \times \Pom$
exchanges. Here the notation corresponds to 
external $\times$ internal $\times$ external exchanges (see
Fig.~\ref{fig:diagrams_4pi_Born}).
Adding $f_{2 \Reg}$ exchange not only enhances the cross section
but also modifies the shape of the distribution.
One can observe now clear enhancements at $y \approx \pm 6$ that
correspond to the external exchanges of $f_{2 \Reg}$ reggeons. This figure
reminds a similar figure for the $p p \to p p \pi^+ \pi^-$ reaction,
where a camel-like distribution was obtained \cite{LS_2pi}.
There the peaks at large rapidities correspond to $f_{2 \Reg}$ reggeon
exchanges. Here (for the $p p \to p p \pi^+ \pi^- \pi^+ \pi^-$ reaction) 
three peaks can be observed. 
In addition, we plot $(\Pom+f_{2 \Reg}) \times f_{2 \Reg} \times (\Pom+f_{2 \Reg})$ 
when in the middle only $f_{2 \Reg}$ is present. 
The cross section is significantly smaller, which means that $\Pom$ in the middle
of the diagram is responsible for the cross section enhancement. 
There is a qualitative hydromechanical analogy in which all outgoing particles 
in diagrams (shown schematically in Fig.~\ref{fig:diagrams_4pi_Born}) are represented as liquid layers 
which move with parallel velocities and protons are the top and bottom layers. 
Then the coupling/friction between layers is given 
by the parameters of the pomeron/reggeon exchanges.
In Fig. \ref{PH_CM_pionsPomeron}, $M_{4\pi}$ distributions for
$(\Pom+f_{2 \Reg}) \times (\Pom+f_{2 \Reg}) \times (\Pom+f_{2 \Reg})$, $\Pom \times \Pom \times \Pom$ and $(\Pom+f_{2 \Reg}) \times f_{2 \Reg} \times (\Pom+f_{2 \Reg})$ are plotted. We can see that at $\Lambda_{off,E}$
 =1.5 GeV and $M_{4\pi} \approx 8 $ GeV the tripple Pomeron exchange contributes $\approx 1/3$ of the cross section reaching $\approx 1/2$ at 30 GeV. These ratios are even smaller for $\Lambda_{off,E}$ = 1 GeV.

In Fig. \ref{PH_CMij} we discuss distribution in dipion invariant mass separately
for the opposite-sign pions (left panel) and for the same-sign pions (right panel). 
To improve statistics and reduce fluctuations the distributions 
for different combinations of indices (34, 56, 36, 45 for the opposite-sign pions 
or 35, 46 for the same-sign pions) 
were averaged in all figures of this type. 
The distributions for the opposite-sign pions
have a large component at low ($M_{\pi \pi} < 3$~GeV) dipion invariant mass, 
similarly to the  dipion mass distribution for the exclusive dipion production 
(see e.g. \cite{LNS2016_2pi}). The distribution for the same-sign pions is clearly broader
than that for the opposite-sign pions and has maximum at larger invariant masses.
This is due to the possible presence of the rapidity gap between the two $\pi^{+}\pi^{-}$ pairs, as illustrated in Figs. \ref{Full_MCentral} and \ref{Atlas_MCentral}, where we plot the differential cross section for the $(+-)(+-) + (-+)(-+)$, $(+-)(-+)+(-+)(-+)$ and $(++)(--)+(--)(++)$ configurations of ordered in rapidity pions as a function of the central rapidity gap width. It is clear that the same sign pion pairs are often formed across the large rapidity gap, what is reflected by the width of their invariant mass distribution Fig. \ref{PH_CMij}. This is also related to the somewhat arbitrary cut-off approach to the region where the Regge formalism (\ref{amplitude_for_pipi_subproces}) does not apply. The higher invariant dipion masses are only weakly
dependent on the cut-off of the low masses
of the two pions across the pomeron/reggeon exchange.
It is not clear to us how to correctly include this region.

The shapes of the distributions in dipion invariant masses only slightly
depend on the value of the cut-off parameter of the off-shell form factor (\ref{off-shell_formfactor}).
The position of the maximum for the same-sign pions at $M_{\pi \pi}
\sim$ 3 GeV seems to be related to the point in $M_{\pi \pi}$ where 
we gradually screen off the Regge amplitude (see Eq.~(\ref{amplitude_for_pipi_subproces})). 
The position of the transition from the Regge to non-Regge physics
have been taken here (somewhat arbitrarily) to be 3 GeV. 
Therefore our predictions are valid above $M_{\pi \pi} >$ 3 GeV. 
What happens below $M_{\pi \pi}$ = 3 GeV is rather a matter of future
measurements. Clearly our approach is not valid in this region
and therefore has no predictive power there.

\begin{figure}[htp]
  \centering
  \includegraphics[width=0.35\textwidth, angle = -90]{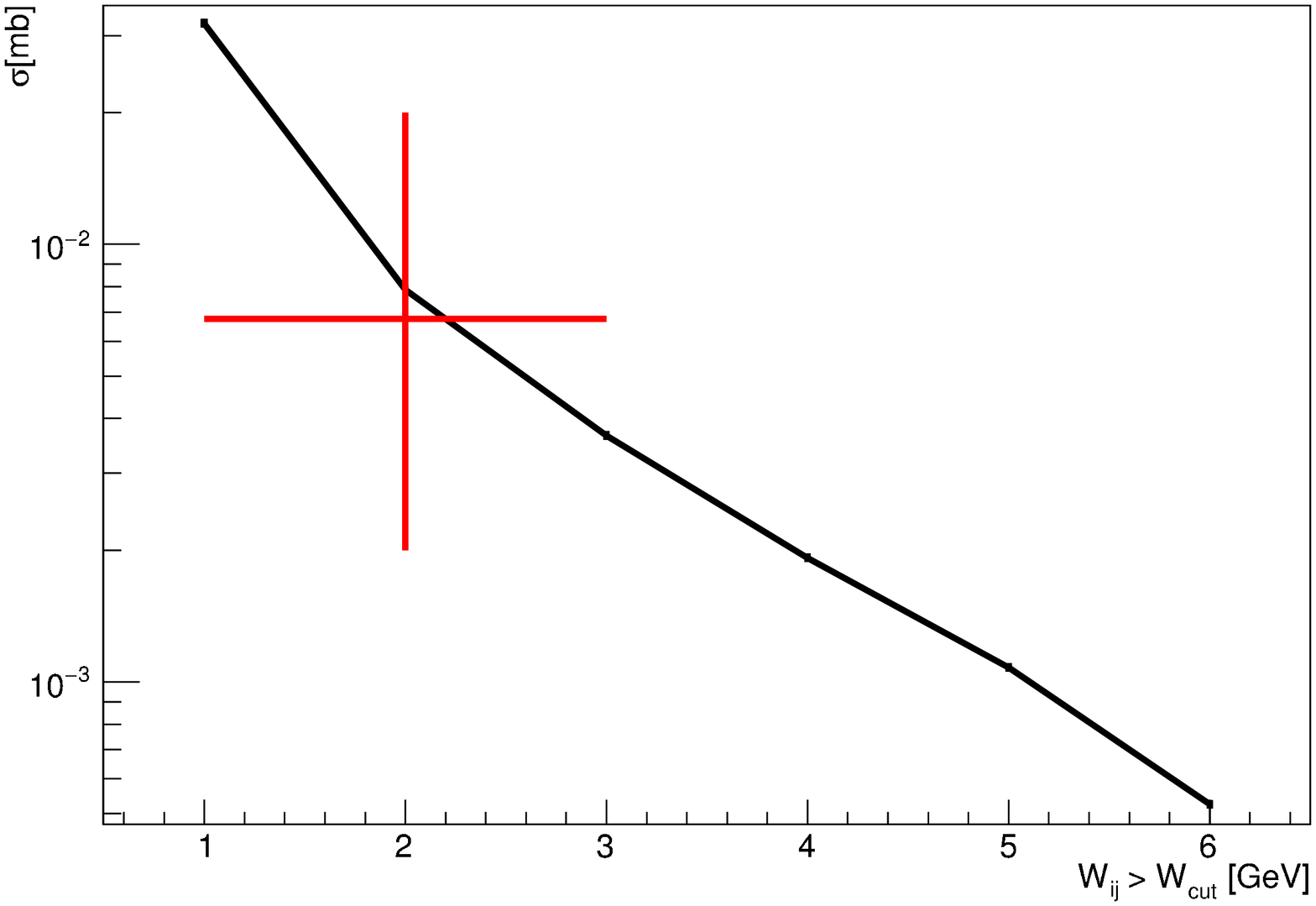}
  \includegraphics[width=0.35\textwidth, angle = -90]{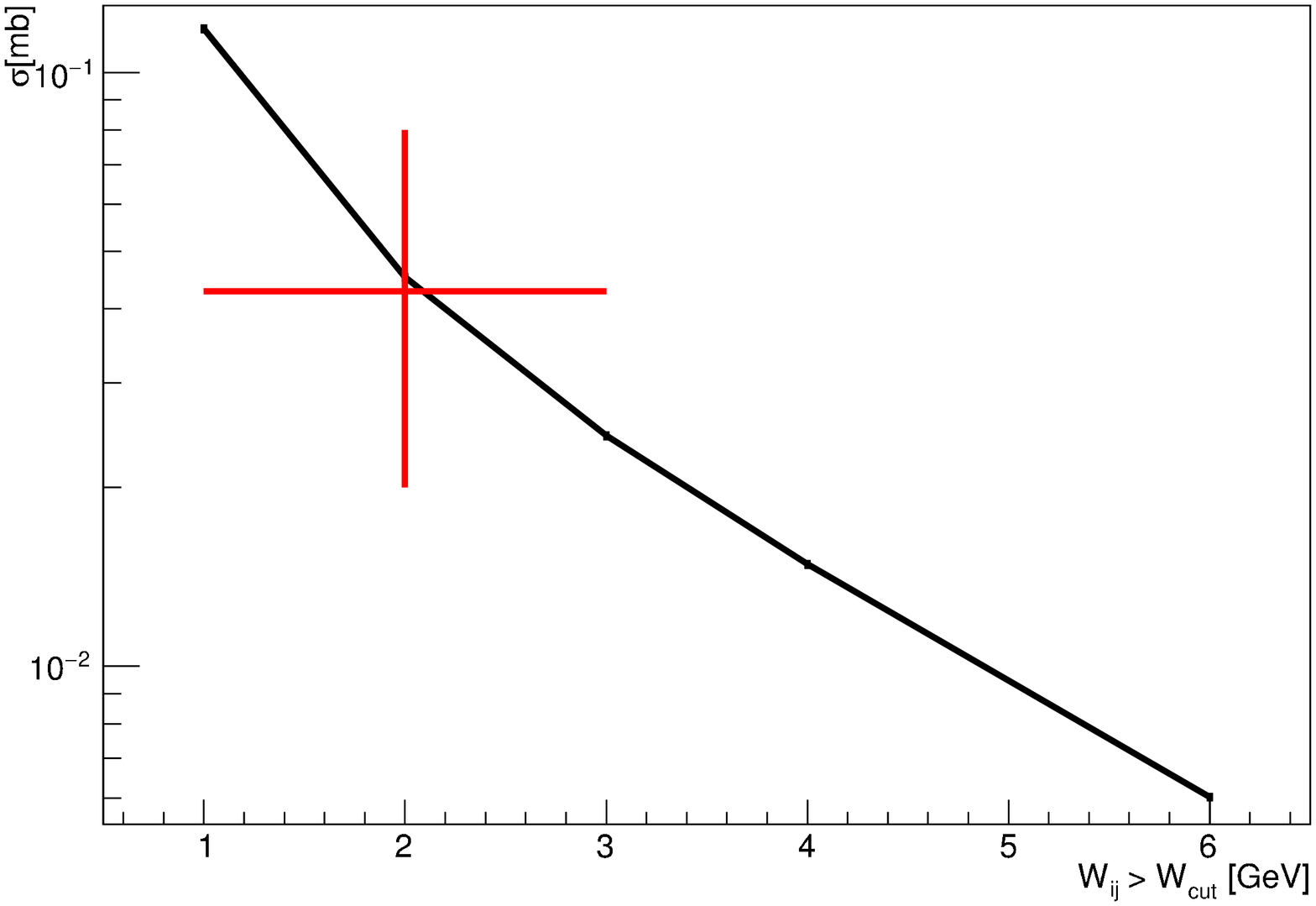}
  \caption{Integrated cross section for different values of
    $W_{cut}$ for sharp (Heaviside-like) cut-off function.       
      The extra (red on line) cross represents cross section for 
      the smooth cut-off function with $W_{0}$ = 2 GeV and $a$ = 0.2 GeV 
      as taken from \cite{LS_2pi}. The left figure is for $\Lambda_{off,E}=1$~GeV and the right one for $\Lambda_{off,E}=1.5$~GeV.}
\label{fig:FullWCut}
\end{figure}

For illustration in Fig.~\ref{fig:FullWCut}
we show dependence of the integrated cross section on the sharp 
cut-off parameter $W_{cut}$ (see Eq.~(\ref{thetaCut})). 
One can observe a power-like dependence of the cross section
as a function of $W_{cut}$.
The extra crosses in the figure show the value of $W_0$ and 
the corresponding cross section 
in the smooth cut-off approach (see Eq.~(\ref{smoothCut})). 
Such a value of $W_0$ was used in the description of 
the $p p \to p p \pi^+ \pi^-$ process (see~\cite{LS_2pi}).
As it is seen the cross section for $W_{0}$ = 2 GeV (smooth cut-off)
is very much the same as the cross section for $W_{cut} \sim 2 - 2.5$ GeV (sharp cut-off).

\begin{figure}[htp]
  \centering
  \includegraphics[width=0.8\textwidth, angle = -90]{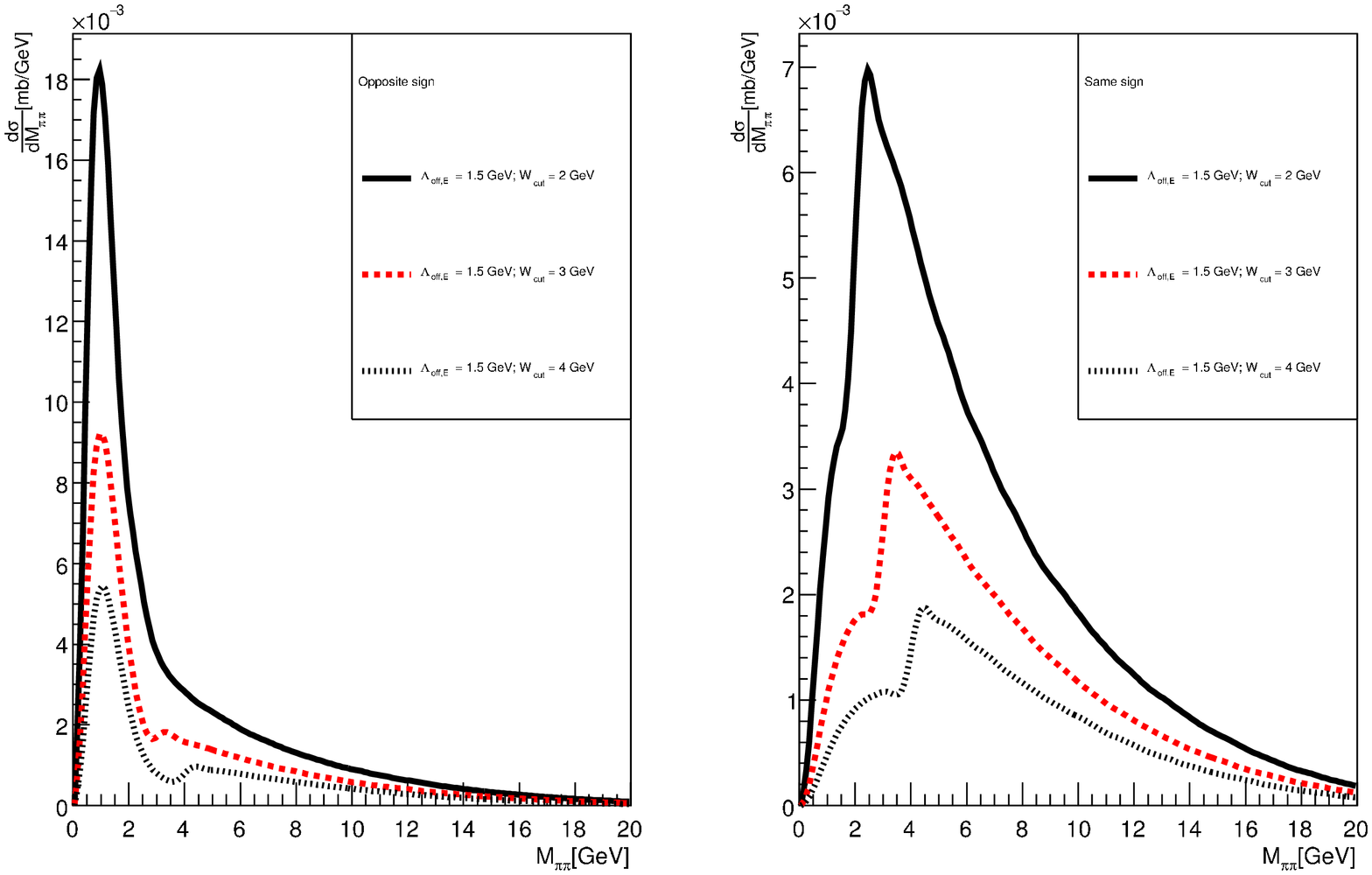}
  \caption{Dipion invariant mass distributions 
          for the opposite-sign (left panel) and the same-sign (right panel)
          pions for different values of $W_{cut}$ for sharp cut-off function (\ref{thetaCut}) and $\Lambda_{off,E}=1.5$~GeV.}
\label{fig:DipDiscontionus}
\end{figure}

Finally, in Fig.~\ref{fig:DipDiscontionus} we show how the choice of $W_{cut}$ (sharp cut-off)
influences the distributions. 
As an example we consider dipion mass spectra, where the effect of cut-off function is the most visible.
In this calculation we fixed $\Lambda_{off,E}$ = 1.5 GeV and we show results for three different values of $W_{cut}$. 
The sharp cut-off leads to characteristic sudden increase of the cross section. 
The large part of the dipion distributions
is only weakly dependent on the value of $W_{cut}$,
but some visible effect survive
(see the location of the dips in Fig.~\ref{fig:DipDiscontionus}). 
This means that the different
dipion subsystems are to some extend correlated.
The study performed here was only to illustrate the possible
uncertainties of our predictions.
However, we believe that our default smooth cut-off is the optimal
choice at present. We feel one should return to the problem when
corresponding experimental data will be available.

\subsection{Results for ATLAS cuts}

In this subsection we present results relevant for the ATLAS 
experimental cuts. The following kinematical conditions are imposed:
\begin{equation}
  |t_{1}|, |t_{2}| < 1 \; {\rm GeV}^{2}\,, \quad
  |y_{\pi}| < 2.5\,, \quad
  p_{t, \pi} > 0.5 \; {\rm GeV}\,.
 \label{ATLASCut1}
\end{equation}

In addition, the mentioned above technical cut $M_{4\pi}<30$ GeV is imposed.

\begin{figure}[htp]
  \centering
  \includegraphics[width=0.35\textwidth, angle = -90]{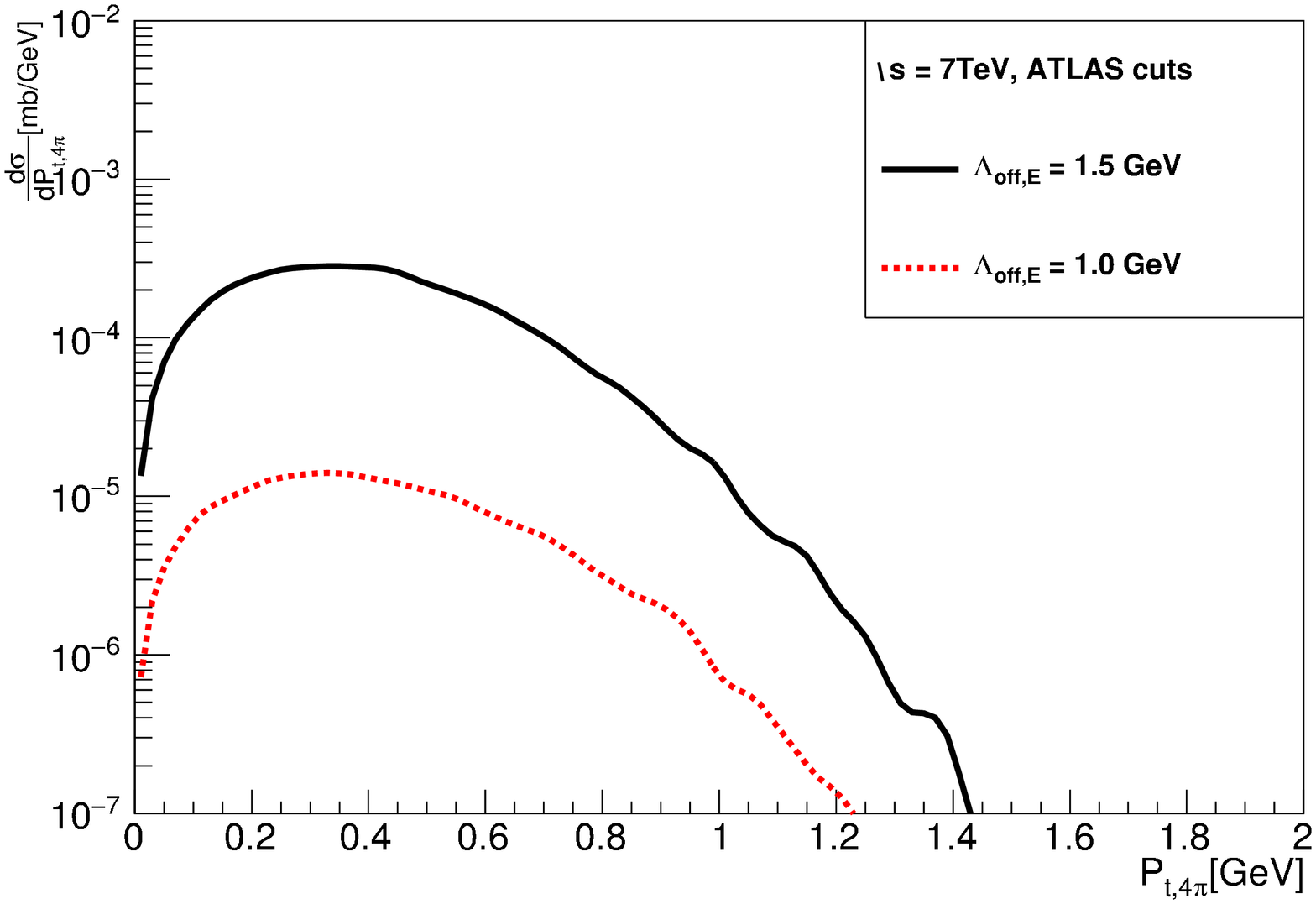}
  \includegraphics[width=0.35\textwidth, angle = -90]{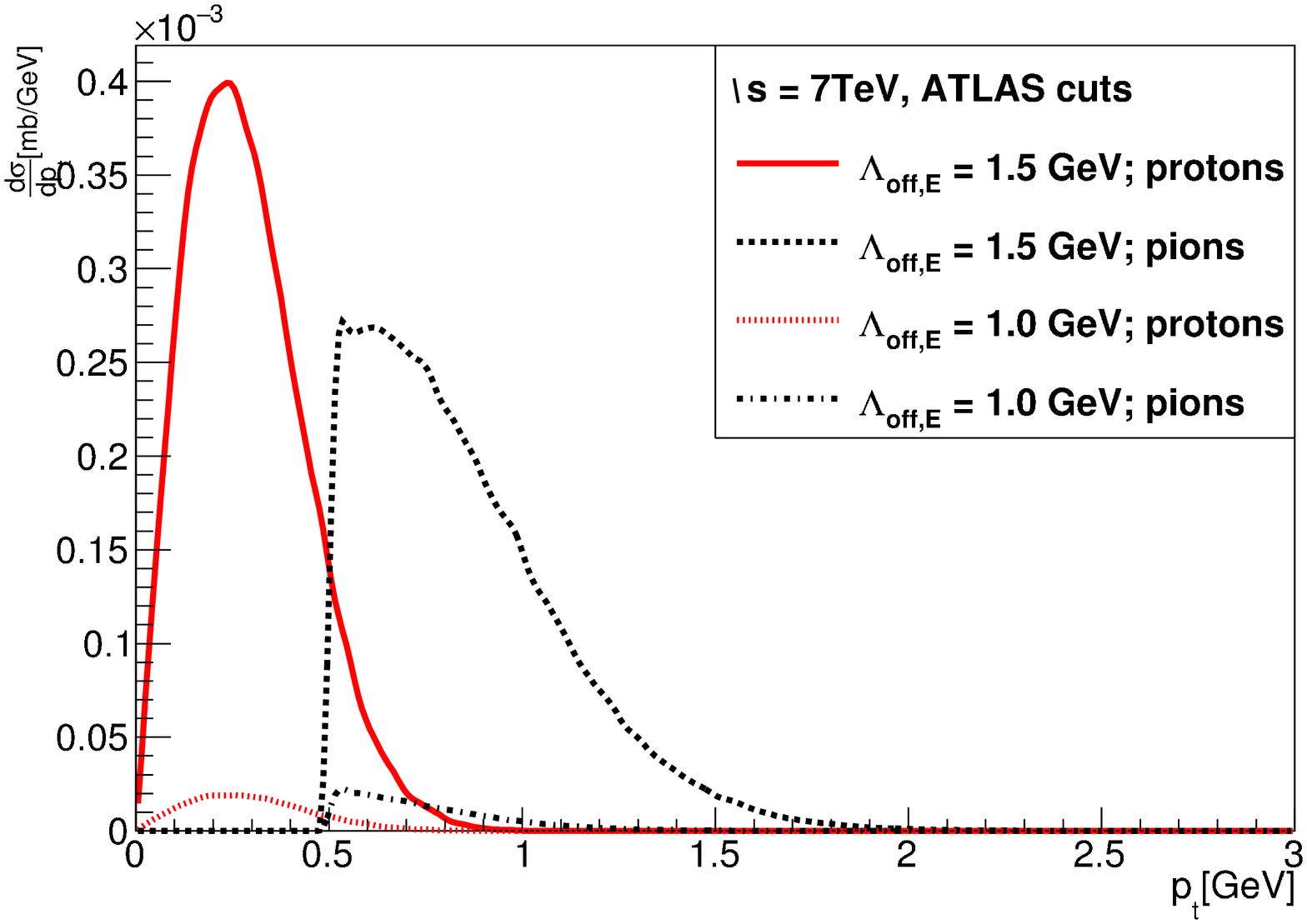}
  \caption{Distribution in transverse momentum of the four-pion system 
          ($P_{t}$) (left panel)
          and for the transverse momenta of individual particles
          (protons or pions) (right panel) with the ATLAS cuts (\ref{ATLASCut1}).}
\label{ATLAS_Pt}
\end{figure}

\begin{figure}[htp]
  \centering
  \includegraphics[width=0.5\textwidth, angle = -90]{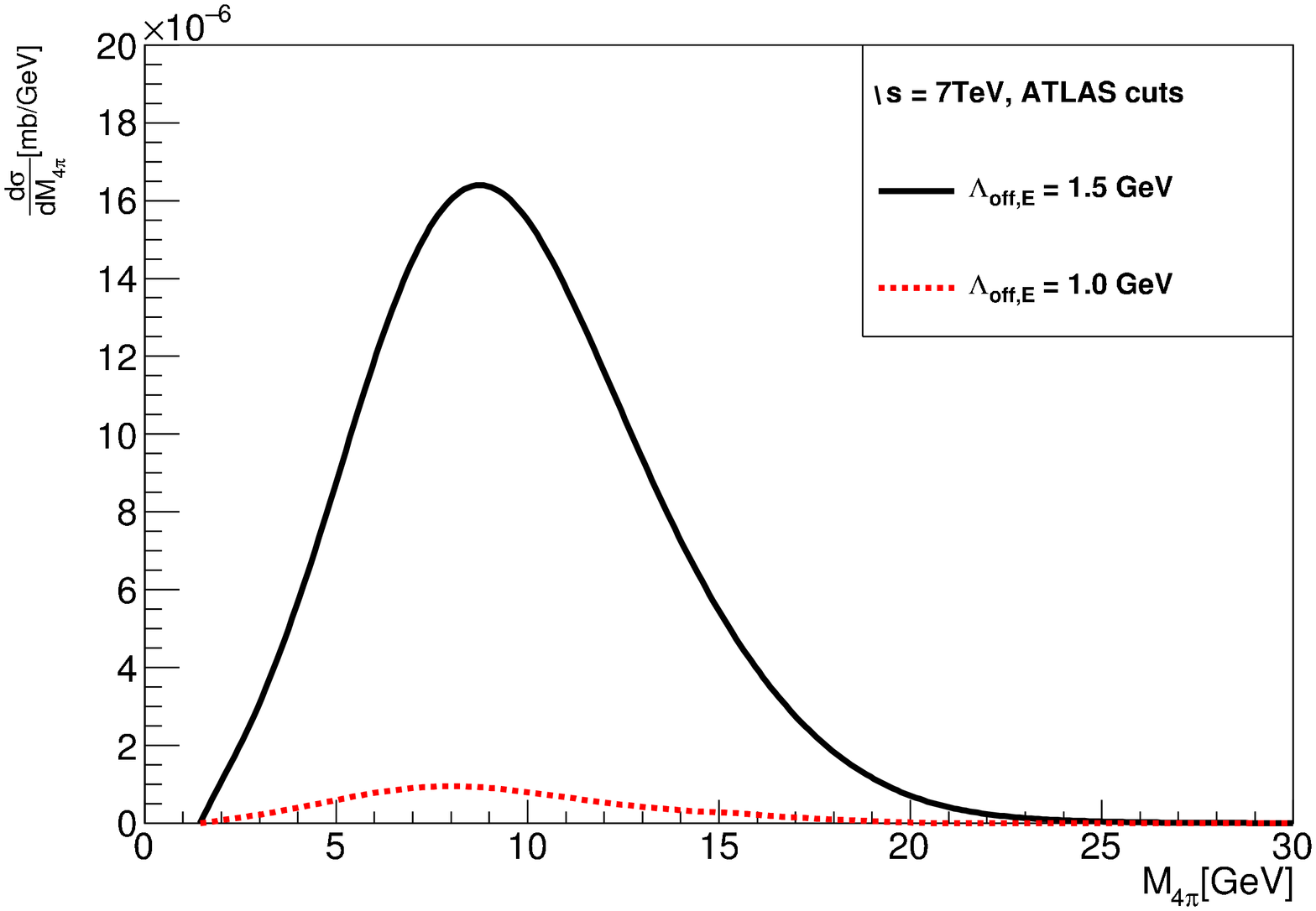}
  \caption{Four-pion invariant mass distribution ($M_{4 \pi}$) 
          with the ATLAS cuts (\ref{ATLASCut1}) for
          $\Lambda_{off,E}=1$~GeV (lower curve) and $\Lambda_{off,E}=1.5$~GeV (upper curve).}
\label{ATLAS_CM}
\end{figure}

\begin{figure}[htp]
  \centering
  \includegraphics[width=0.5\textwidth, angle = -90]{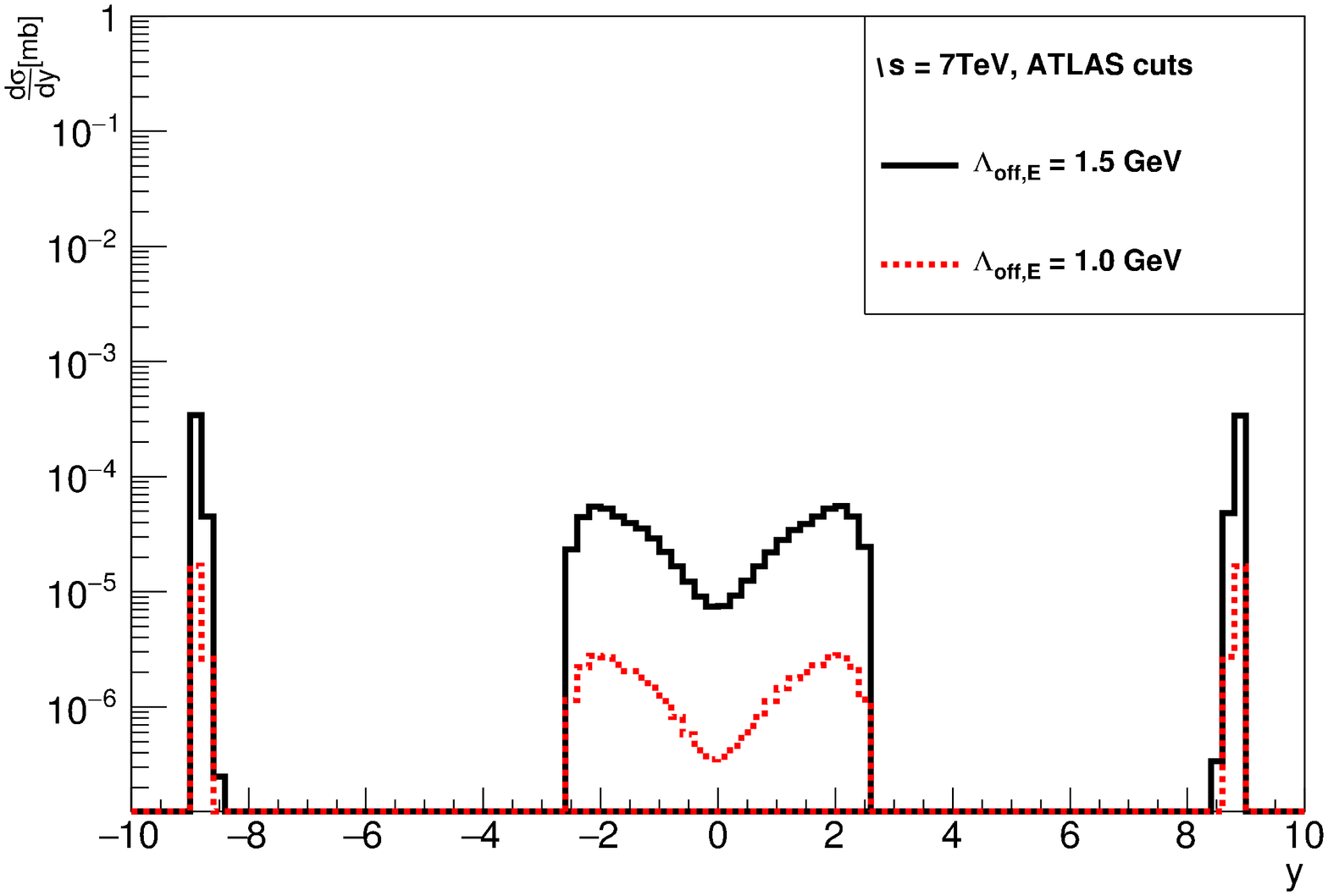}
  \caption{Distribution in rapidity of pions and protons for the ATLAS cuts (\ref{ATLASCut1}).}
\label{ATLAS_Y}
\end{figure}

\begin{figure}[htp]
  \centering
  \includegraphics[width=0.8\textwidth, angle = -90]{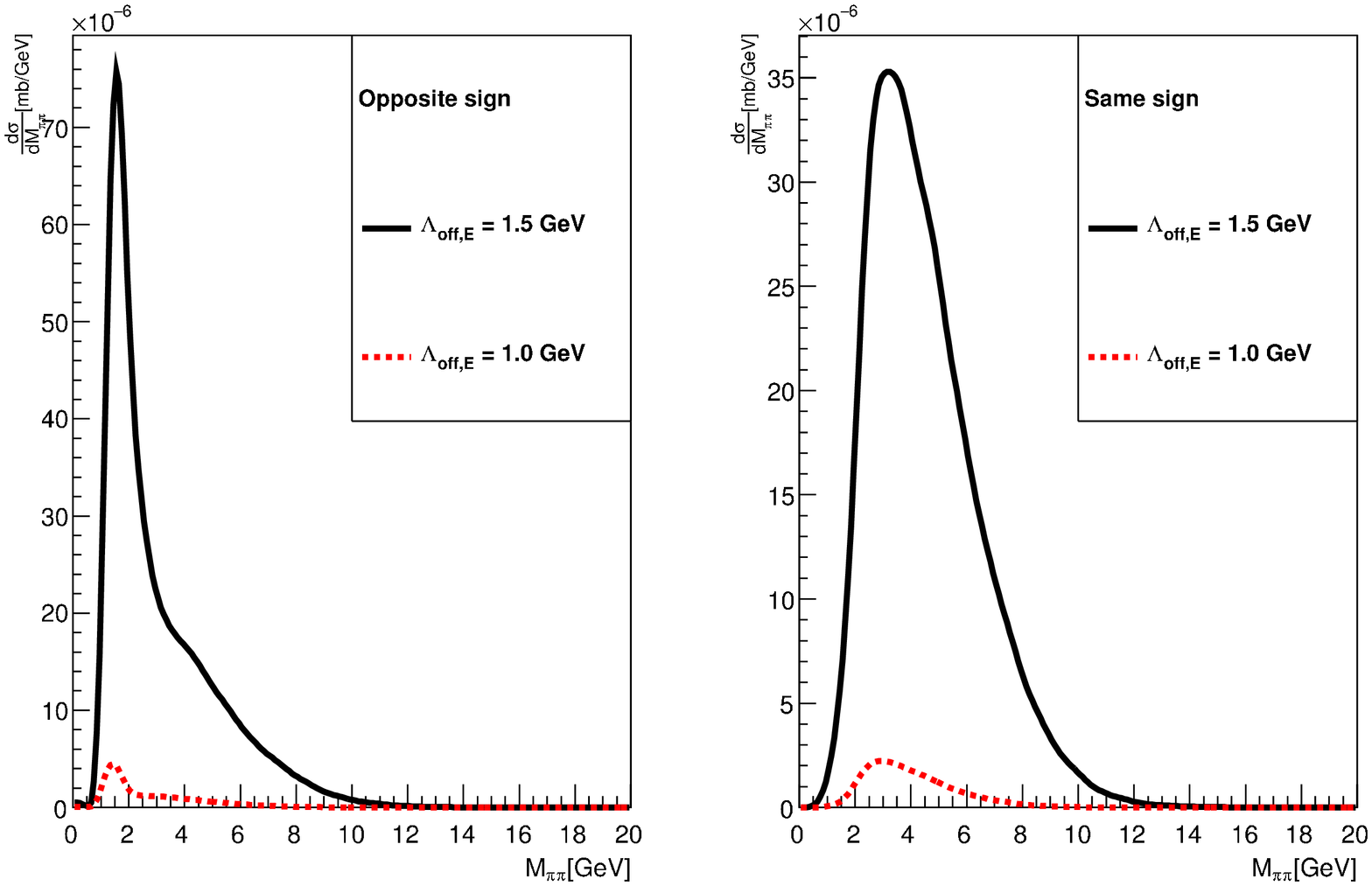}
  \caption{Dipion invariant mass distribution for the
    opposite-sign (left panel) and same-sign (right panel) pions
    with the ATLAS cuts (\ref{ATLASCut1}) for different values of $\Lambda_{off,E}$.}
\label{ATLAS_CMij}
\end{figure}

The corresponding integrated cross sections for different values of 
the cut-off parameter are collected in Tab.~\ref{tab:CrossSectionATLAS}.
In this case the dependence on $\Lambda_{off,E}$ is even stronger than for
the full phase space case. 
This means that precise prediction of the cross section is not simple.
\begin{table}[!htb]
  \caption{Integrated cross sections in nb with the ATLAS cuts (\ref{ATLASCut1})
           for different values of the cut-off parameter $\Lambda_{off,E}$.
           No absorption effects are included here.}
\centering
  \begin{tabular}{| l | c | c | }
    \hline
           & $\Lambda_{off,E}$ [GeV] & $\sigma$ [nb]  \\ \hline
    ATLAS & 1.0 & \;\;\;\,6.91 \\ \hline
    ATLAS & 1.5 & 141.43  \\ \hline
  \end{tabular}
\label{tab:CrossSectionATLAS}
\end{table}
%
Similarly as for the full phase space case we present several
differential distributions in Figs.~\ref{ATLAS_Pt}, \ref{ATLAS_CM}, \ref{ATLAS_Y}, \ref{ATLAS_CMij}. 

The transverse momentum distribution of the four-pion system is shown
in Fig.~\ref{ATLAS_Pt}. The shape of the distribution is practically
the same as for the full phase space case.
It should be noted that the distribution of the  longitudinal momentum of the four-pion system after rapidity cuts related to the ATLAS central tracker acceptance (right side of Fig. \ref{pz_distribution}) is very narrow in comparison to full phase space histogram shown in the same figure on the left side.
The four-pion invariant mass distribution extends from 
$M_{4 \pi} \approx 3$~GeV to the upper cut at $M_{4 \pi} = 30$ GeV.
This means that the ATLAS experiment has a potential 
to investigate the discussed here mechanism.

The rapidity distributions of pions (middle bump) and protons 
(external peaks) are shown in Fig.~\ref{ATLAS_Y}.
Here the rapidity coverage of the main tracker is clearly visible.
The rapidity gaps between protons with $y \approx \pm$ 9 and pions 
are now set by the experimental cuts and are bigger than 4.5 rapidity units. 
But we have to assure, in addition, the existence
of rapidity gap within the four-pion system confined now to $|y_{\pi}| <  2.5$. 
Then the maximal rapidity gap is clearly confined from above to only five rapidity units.

In Fig.~\ref{ATLAS_CMij} we show dipion
invariant mass distribution for the opposite-sign (left panel) and the same-sign (right panel) pions for two different values of $\Lambda_{off,E}$.

One major weakness of the discussed model is a rather simplistic treatment of the low dipion and $\pi p$ invariant masses i.e. non-Regge region. These region can be removed from the data imposing additional cut  $M_{ij} > M_{ij, cut} \approx$~2 -- 4~GeV.  
Imposing such cuts leads to the cross sections 
collected in Tab.~\ref{tab:CrossSection_extra_cuts}.
The rate of the reduction of cross section depends on the value of the cut-off parameter and the way how amplitudes are modified
in the difficult to control non-Regge region.
\begin{table}[!htb]  
  \caption{The integrated cross sections in nb for the ATLAS cuts (\ref{ATLASCut1}) with 
the extra limitations on $\pi\pi$ and $p\pi$ invariant masses.
The columns 'Smooth' show the resulting cross sections for the cut-off function of (\ref{smoothCut}). The columns 'Sharp' show results obtained for the cut-off function of (\ref{thetaCut}) with $W_{cut}$ = 2 GeV. No absorption effects are included here.}
\centering

  \begin{tabular}{| l | c | c | c | c | }
  \hline
&  
\multicolumn{2}{|c|}{$\Lambda_{off,E}$ = 1.0 GeV}  &   
\multicolumn{2}{|c|}{$\Lambda_{off,E}$ = 1.5 GeV} \\ \hline
             &   Smooth   &  Sharp        & Smooth      &  Sharp \\ \hline
no extra cut on $M_{ij}$ &    7.35     &  6.91  &    148.83       &  141.43  \\
$M_{ij, cut}  = 2$ GeV          &    7.35     &  6.90  &    146.92       &   141.33   \\
$M_{ij, cut} = 3$ GeV          &    6.66     &  6.31  &    138.79       &   134.10   \\
$M_{ij, cut}  =4$ GeV          &    5.15     &  4.82  &    116.54       &   113.73    \\
\hline
  \end{tabular}
\label{tab:CrossSection_extra_cuts}
\end{table}

\begin{figure}[htp]
  \centering
\includegraphics[width=0.7\textwidth, angle = -90]{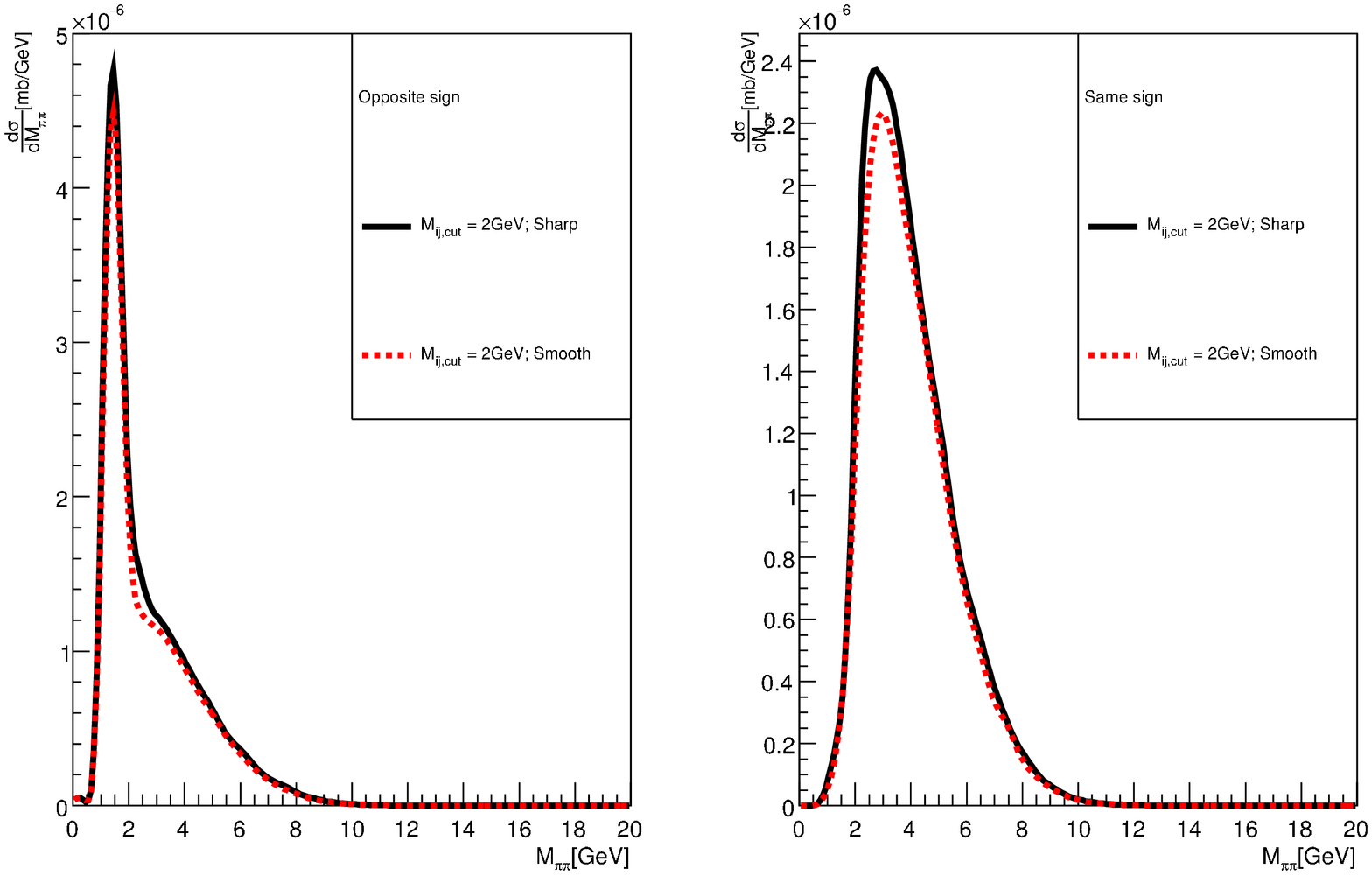}
\includegraphics[width=0.7\textwidth, angle = -90]{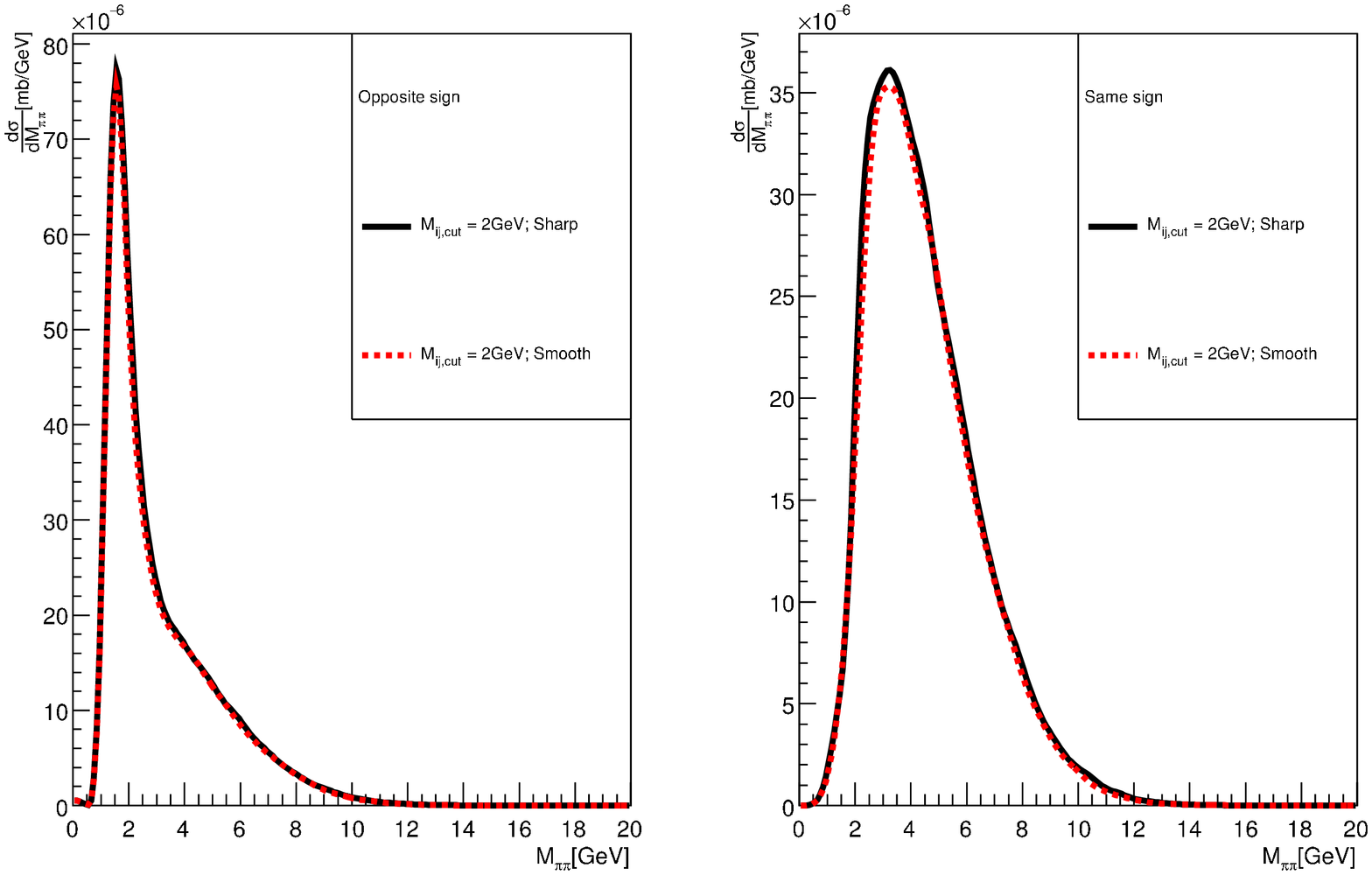}
  \caption{Dipion invariant mass distributions for the
    opposite-sign (left panel) and the same-sign (right panel) pions
    for the ATLAS experimental cuts (\ref{ATLASCut1}) with the extra limitations on $M_{ij}>M_{ij,cut}$. The upper plot is for $\Lambda_{off,E}=1$ GeV and the lower one for $\Lambda_{off,E}=1.5$ GeV. 'Smooth' means cut-off (\ref{smoothCut}) with $W_{0}=2$ GeV and 'Sharp' means cut-off (\ref{thetaCut}) 
with $W_{cut} = 2$~GeV.}
\label{ATLAS_CMijExperimentalCutsFull}
\end{figure}

These experimental cuts remove influence of the details of the cut-off function on $M_{\pi\pi}$ plots, see Fig.~\ref{ATLAS_CMijExperimentalCutsFull}.
These additional cuts ($M_{ij.cut}=2$ GeV) practically do not change the distributions shown in Fig. \ref{ATLAS_CMij}. The same can be seen by comparison of first two rows of Tab. \ref{tab:CrossSection_extra_cuts}.

\subsection{Results for ALICE cuts}

For the ALICE experiment we take the following cuts:
\begin{equation}
  p_{t,p} < 2 \; {\rm GeV}\,, \quad
  p_{t,\pi} > 0.17 \; {\rm GeV}\,, \quad
  |\eta_{\pi}| < 0.9\,,
 \label{ALICECut1}
\end{equation}
and the technical $M_{4\pi}<30$ GeV cut.
Corresponding numerical values for the integrated cross sections 
are presented in Tab.~\ref{tab:CrossSectionALICE}.
They are rather small compared to the ATLAS case.
\begin{table}[!htb]
  \caption{Integrated cross sections in pb for the ALICE cuts (\ref{ALICECut1}) for the smooth cut-off function and
           for different values of $\Lambda_{off,E}$.
           No absorption effects are included here.}
\centering
  \begin{tabular}{| l | c | c | }
    \hline
                 & $\Lambda_{off,E}$ [GeV] &   $\sigma$ [pb]      \\ \hline
    ALICE       &    1.0           &   $\;\,4.2$       \\ \hline
    ALICE       &    1.5           &   $ 37.7$        \\ \hline
  \end{tabular}
  \label{tab:CrossSectionALICE}
\end{table}

\begin{figure}[htp]
  \centering
\includegraphics[width=0.35\textwidth, angle = -90]{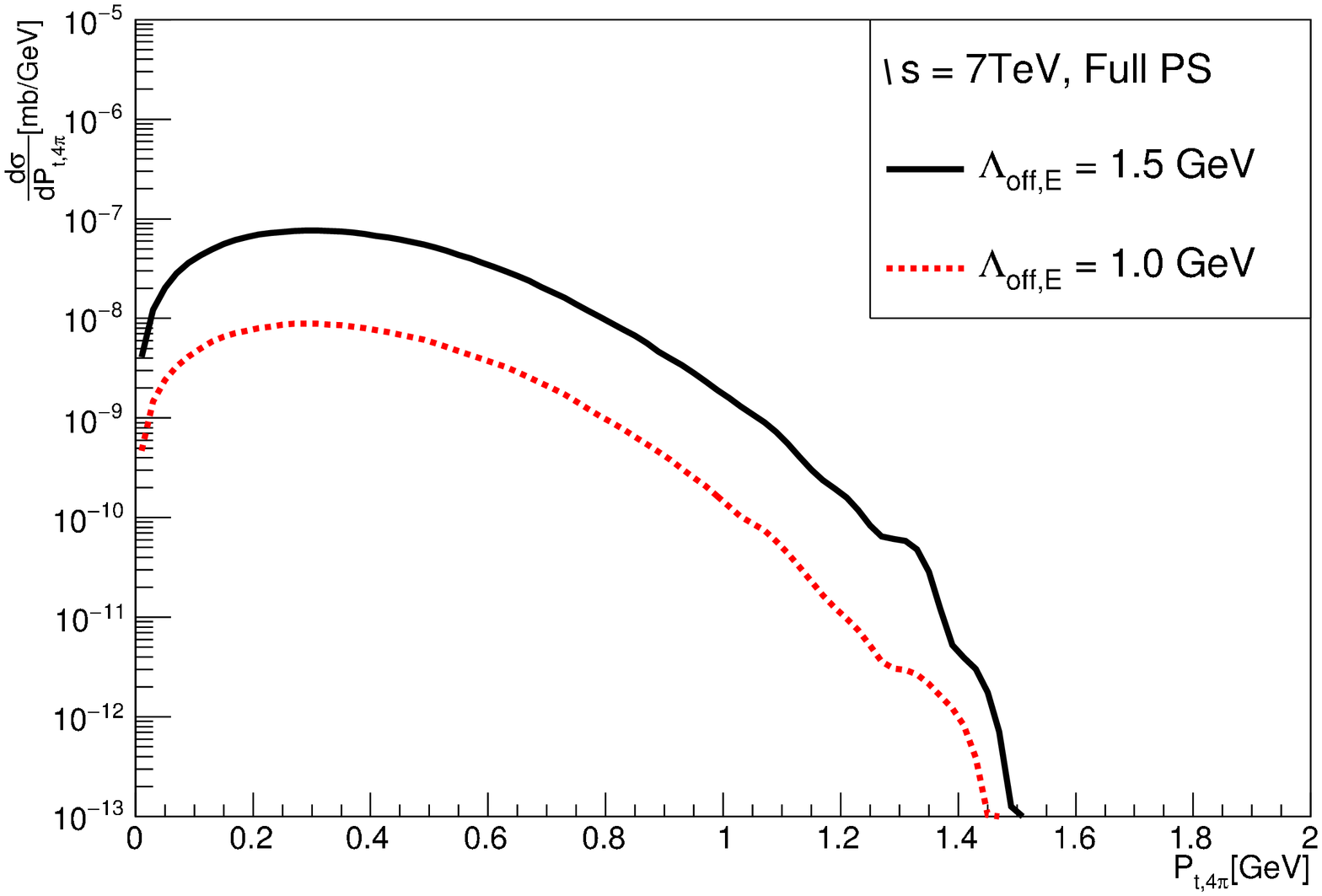}
\includegraphics[width=0.35\textwidth, angle = -90]{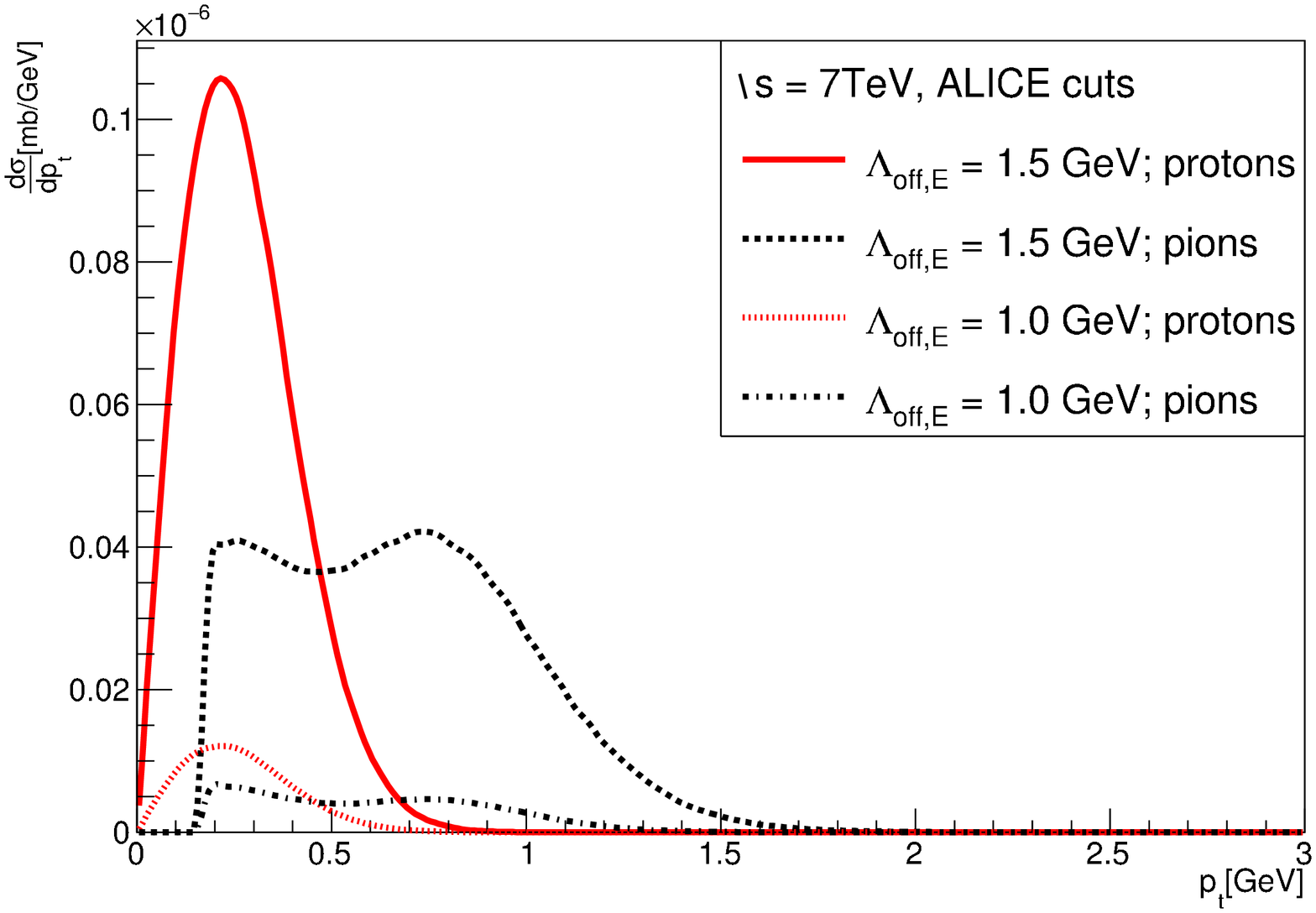}
  \caption{Distribution in transverse momentum of the $4 \pi$ system (left panel) 
    and in transverse momenta of individual pions or protons (right panel)
    for the ALICE cuts (\ref{ALICECut1}) and for the two values of $\Lambda_{off,E}$ specified in the figure legends.}
  \label{ALICE_Pt}
\end{figure}

Several differential distributions are presented below 
in Figs. \ref{ALICE_CM}, \ref{ALICE_Y} and \ref{ALICE_CMij}. 
The distribution in transverse momentum of the four-pion system
(Fig. \ref{ALICE_Pt}) is here very similar as those for the full phase space 
and for the ATLAS cuts. In contrast, the distribution in four-pion 
invariant mass drops faster than its counterpart for the ATLAS case. 
The irregular structures are due to narrow
rapidity coverage of the ALICE detector and/or
the cuts on each exchange of the pomeron or reggeon. 
The cross section for the triple Regge mechanism is for the ALICE fiducial volume very small, see Tab. \ref{tab:CrossSectionALICE}. In addition, other mechanisms (see \cite{LNS2016_4pi}) may be important in this region.
We conclude that the ALICE detector is not well suited for the studies of processes with three pomeron/reggeon exchanges.

\begin{figure}[htp]
  \centering
  \includegraphics[width=0.5\textwidth, angle = -90]{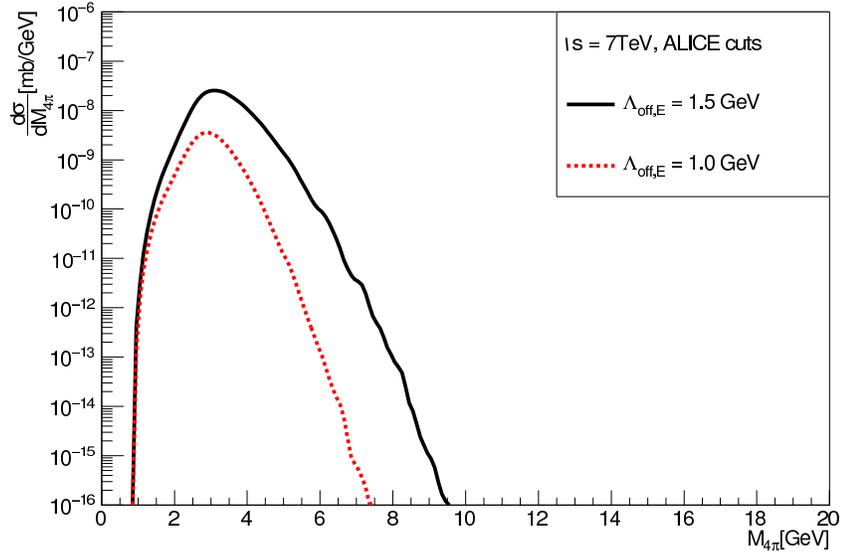}
  \caption{Four-pion invariant mass distribution for the ALICE cuts (\ref{ALICECut1}) 
  and for $\Lambda_{off,E} = 1$~GeV (lower curve) 
  and $\Lambda_{off,E} = 1.5$~GeV (upper curve).}
  \label{ALICE_CM}
\end{figure}

\begin{figure}[htp]
  \centering
  \includegraphics[width=0.5\textwidth, angle = -90]{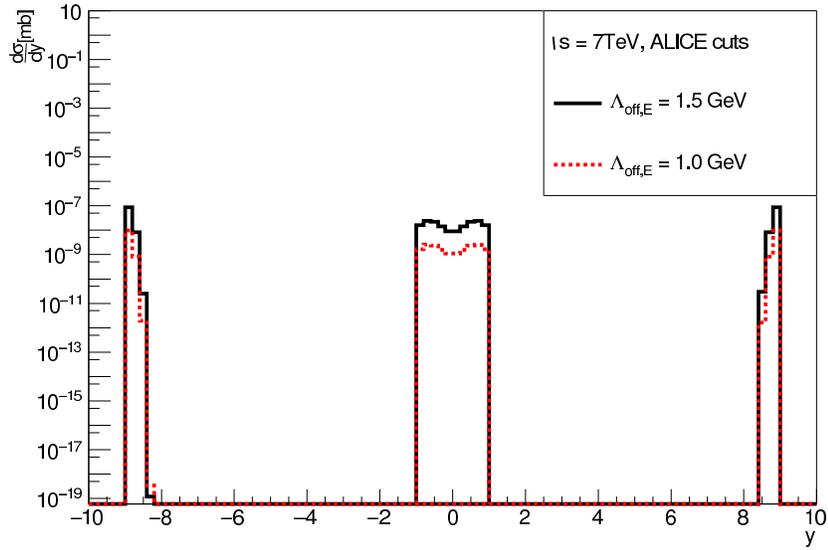}
  \caption{Distribution in rapidity of pions (middle bump) and of protons (external peaks) for the ALICE cuts
           for two different values of $\Lambda_{off,E}$.}
  \label{ALICE_Y}
\end{figure}

\begin{figure}[htp]
  \centering
  \includegraphics[width=0.8\textwidth, angle = -90]{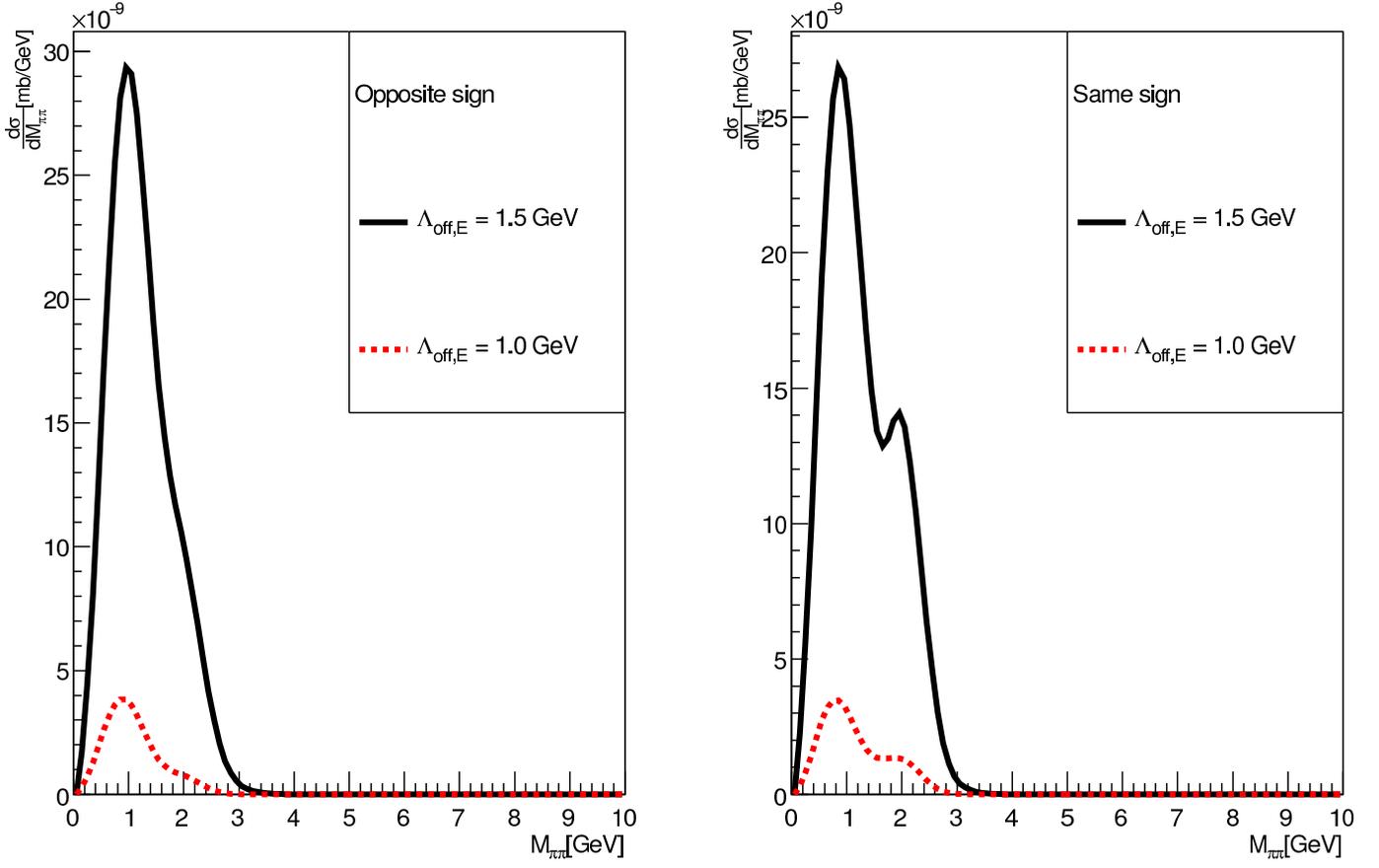}
  \caption{Dipion invariant mass distribution 
  for the opposite  sign (left panel) and the same sign (right panel) pions for the ALICE cuts (\ref{ALICECut1}).}
  \label{ALICE_CMij}
\end{figure}
\section{Discussion of some additional aspects of the triple-Regge exchange model} 
Here  we discuss in more detail some aspects of the model only mentioned in the previous section.

\subsection{$M_{4\pi}$ distribution and energy transfer to the $4\pi$ system}

Here we wish to discuss some kinematic properties of the $4\pi$ system. In particular, we wish to understand how much energy can be transferred to the four-pion system.
In our opinion, this is determined by the fact that the scattered protons 
take almost all energy leaving only a small amount of available energy which is distributed 
among the all centrally produced final pions. This can be traced back
to specific propagators of pomerons/reggeons that couple to protons.
In this subsection we shall try to justify the hypothetical statement.

\begin{figure}[htp]
  \centering
  \includegraphics[width=0.35\textwidth, angle = -90]{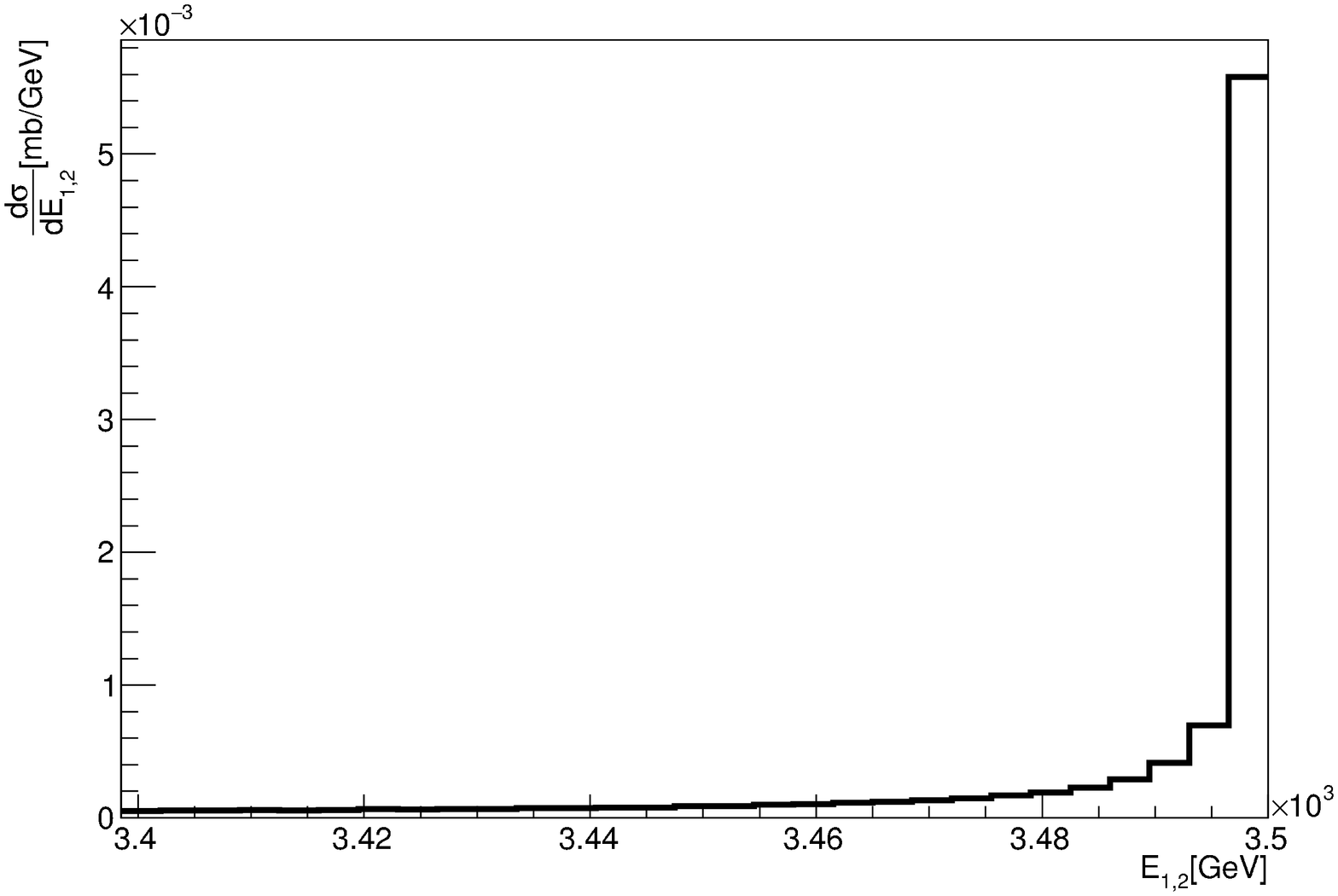}
  \includegraphics[width=0.35\textwidth, angle = -90]{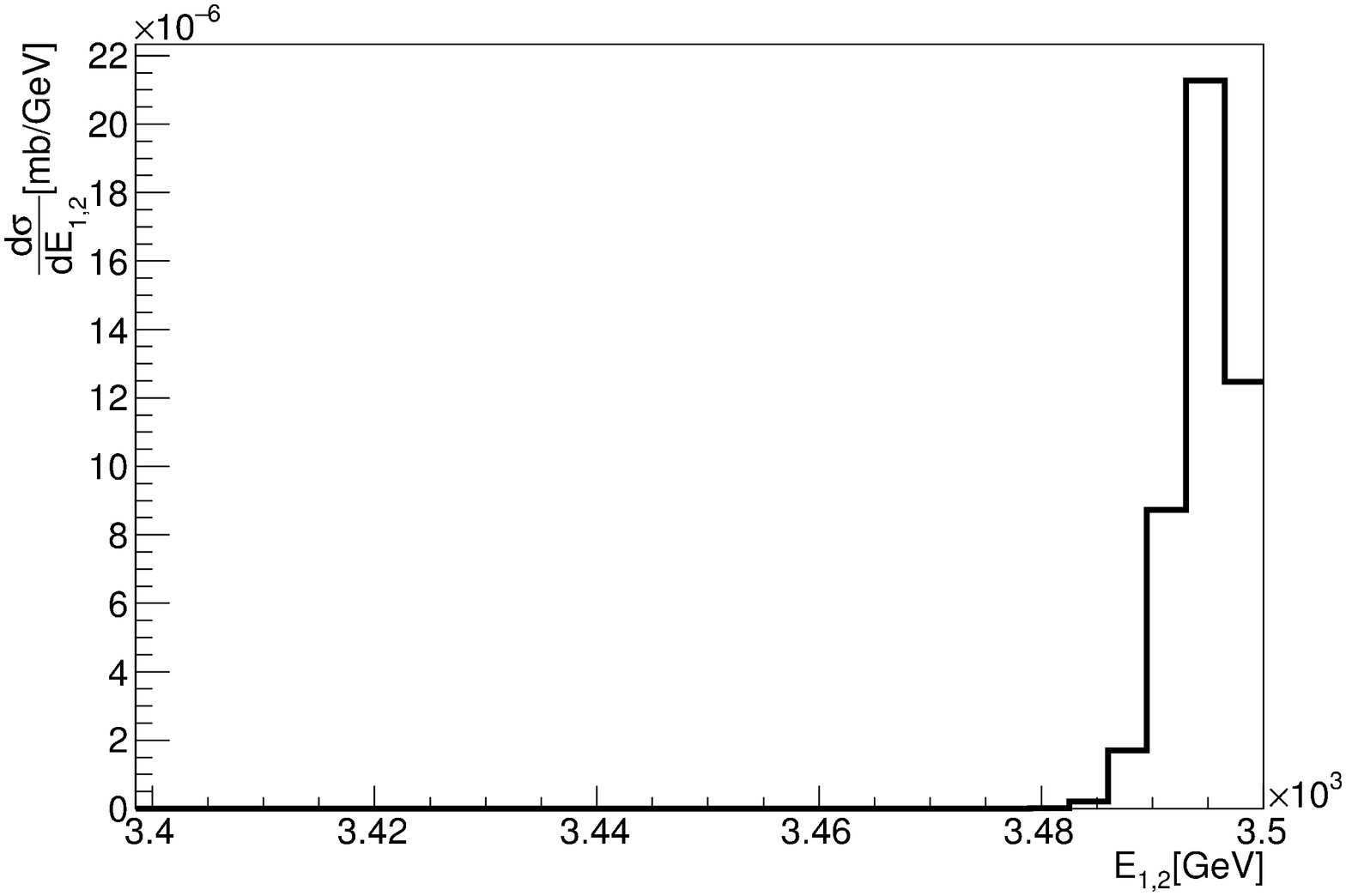}
  \caption{Proton energy distributions for the full phase space (left) 
  and for the ATLAS cuts (\ref{ATLASCut1}) (right).
  Here we take $\Lambda_{off,E}=1.5$~GeV as an example.}
  \label{E1E2_distribution}
\end{figure}

In order to better understand this effect we first plot distribution 
in energy of one of outgoing protons in Fig.~\ref{E1E2_distribution}. 
The distribution quickly drops towards energies smaller than
$\sqrt{s}/2$ which in our example is 3.5 TeV. 
For the case of the ATLAS cuts (\ref{ATLASCut1}) the drop is much faster.

\begin{figure}[htp]
  \centering
  \includegraphics[width=0.35\textwidth, angle = -90]{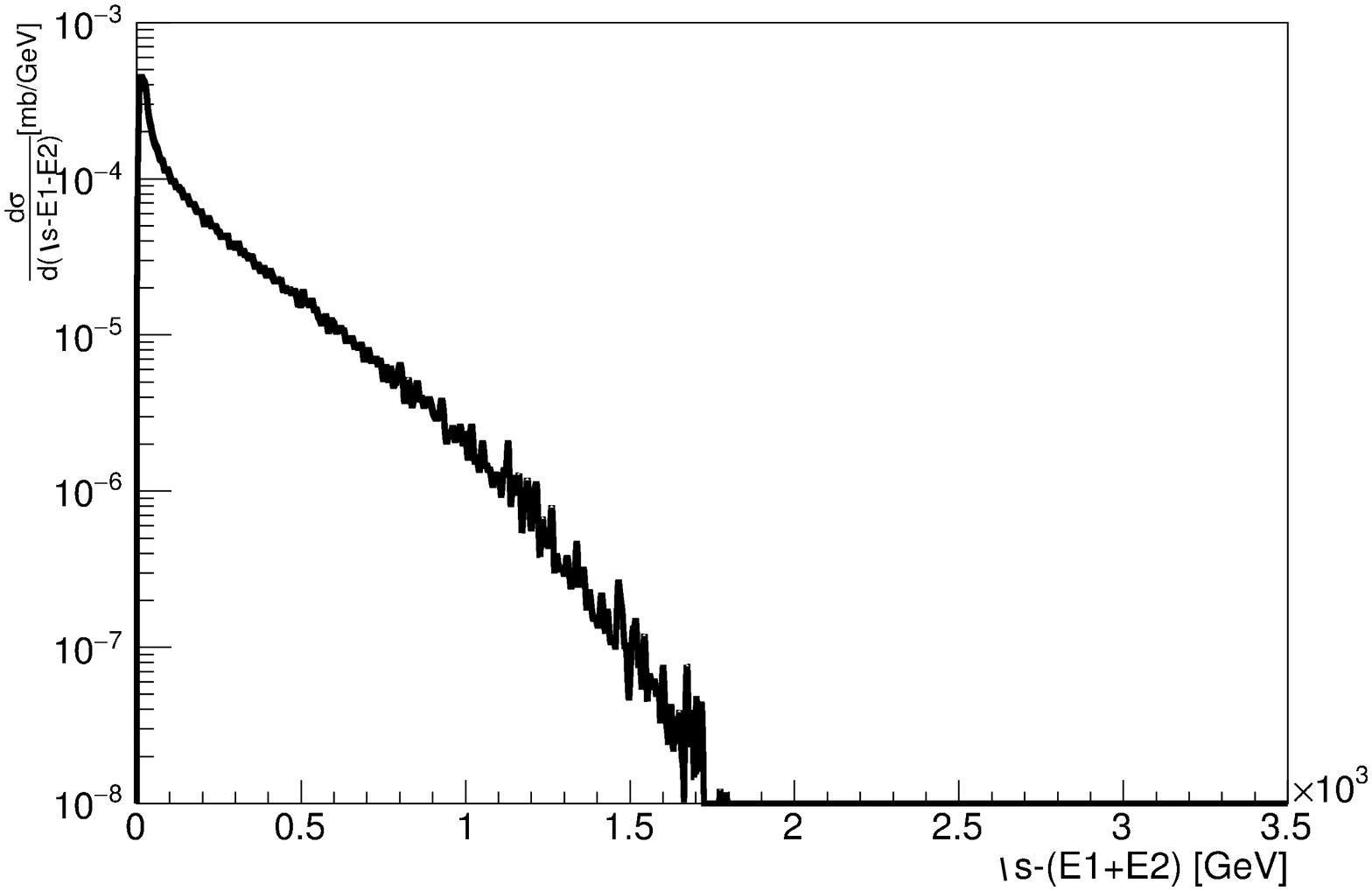}
  \includegraphics[width=0.35\textwidth, angle = -90]{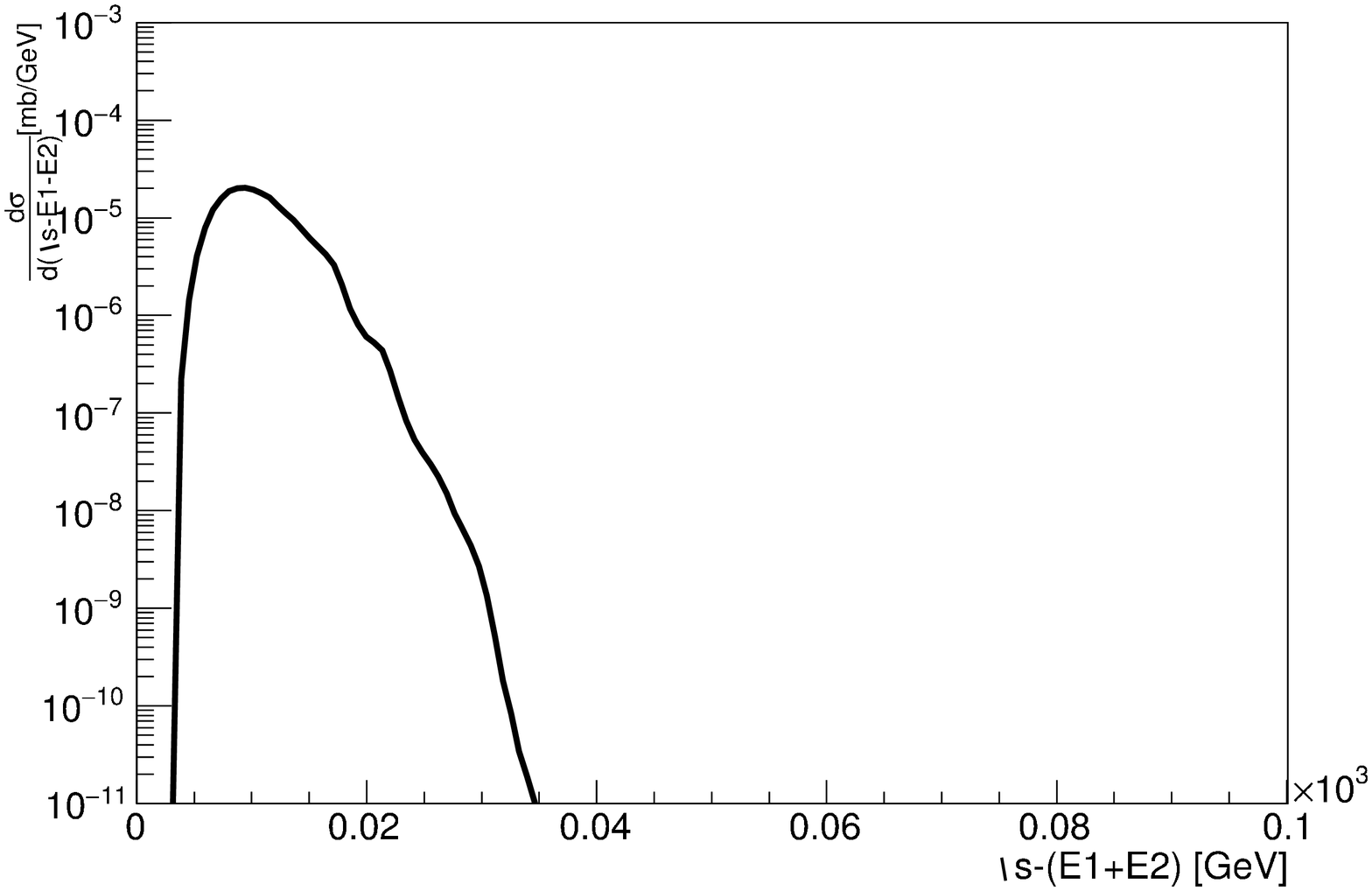}
  \caption{Distributions in the energy left for pions for the full phase space (left) 
  and for the ATLAS cuts (\ref{ATLASCut1}) (right).
  Here we take $\Lambda_{off,E}=1.5$~GeV. For the full phase space case there is visible a sharp break above $1500$~GeV indicating numerical accuracy limit.}
  \label{TECME1E2_distribution}
\end{figure}

Summing energies of both outgoing protons we obtain 
total energy taken by protons. In Fig.~\ref{TECME1E2_distribution} we show distribution 
in the energy left for pions, which is the total energy minus the energy taken away by protons. 
The energies left for pions are much smaller than those taken away by
protons. Kinematics dictates the following inequality:
\begin{equation}
M_{4 \pi} < \sqrt{s} - E_1 - E_2 \,.
\label{inequality}
\end{equation}

For our process the dynamics imposes much severe cuts so it means the restriction (\ref{inequality}) is not really important.

In general, the whole four-pion system does not need to be at rest in the overall
centre of mass system. Let us define the quantity:
\begin{equation}
P_{4 \pi, z} = p_{3,z} + p_{4,z} + p_{5,z} + p_{6,z} \; .
\label{pz}
\end{equation}

\begin{figure}[htp]
  \centering
  \includegraphics[width=0.35\textwidth, angle = -90]{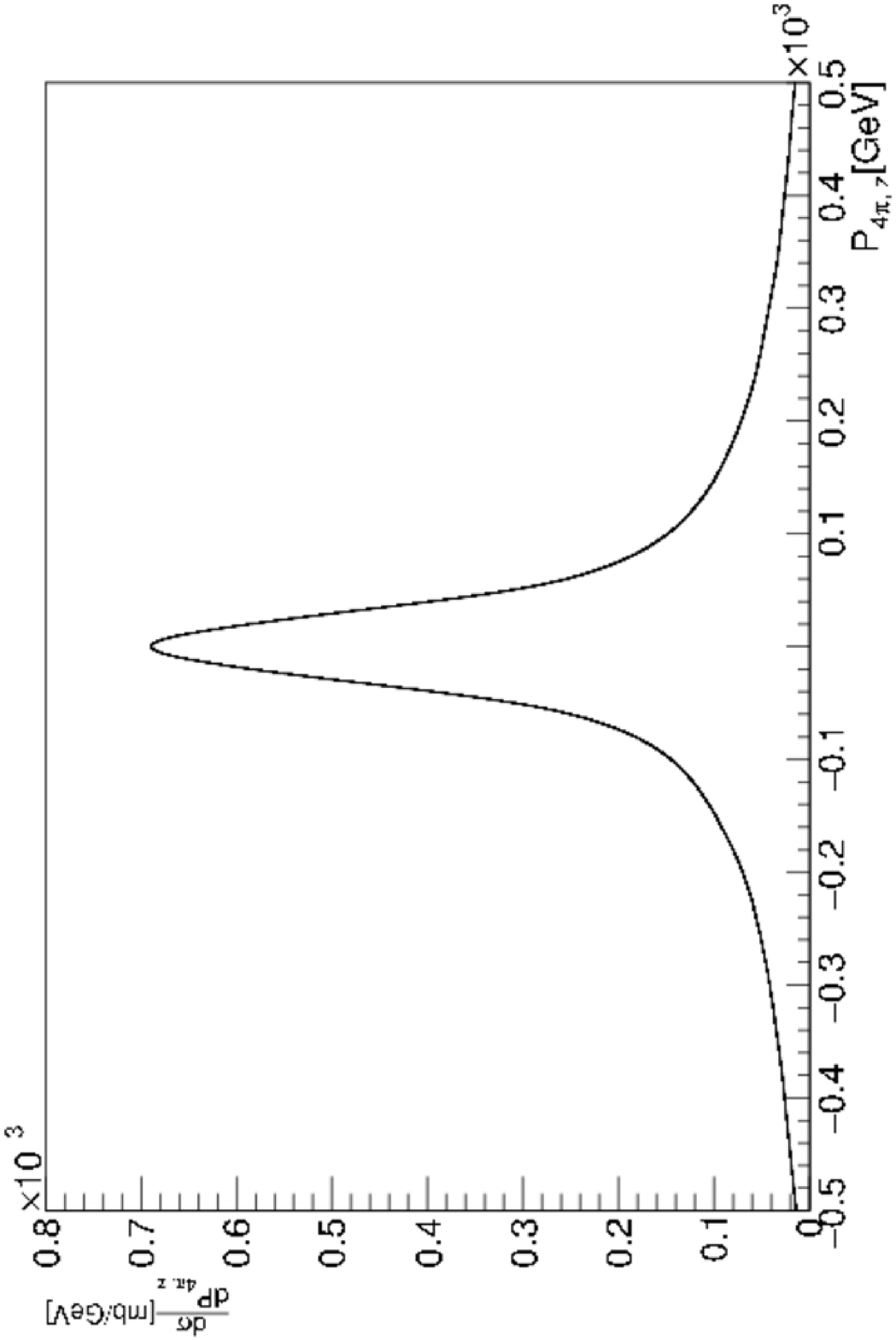}
  \includegraphics[width=0.35\textwidth, angle = -90]{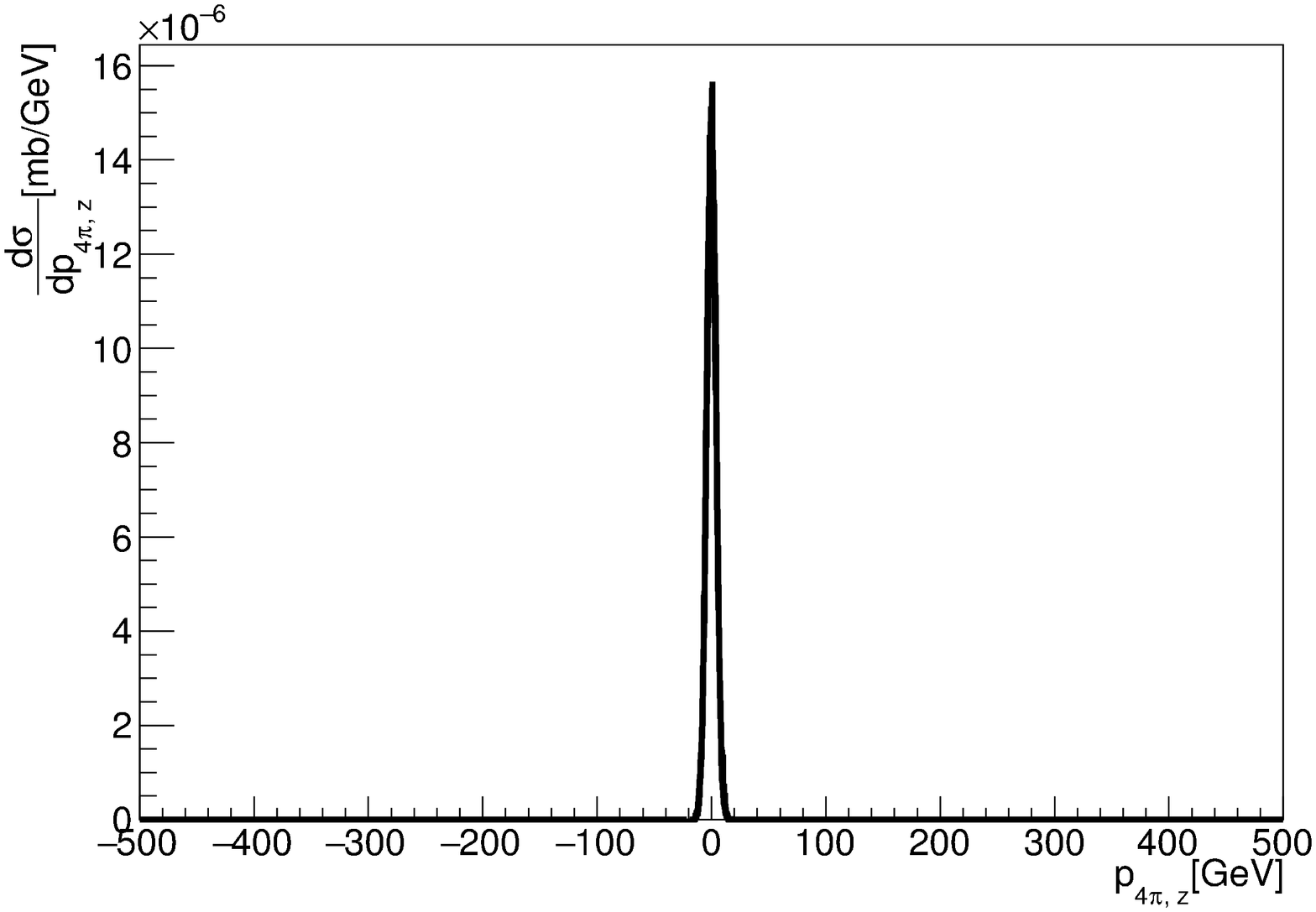}
  \caption{Distributions in the longitudinal momentum of the pion center
  of mass for the full phase space (left) and for the ATLAS cuts (\ref{ATLASCut1}) (right).
  Here we take $\Lambda_{off,E}=1.5$~GeV as an example.}
  \label{pz_distribution}
\end{figure}

In Fig.~\ref{pz_distribution} we show distribution of the $P_{4 \pi, z}$ variable. We observe much narrower distribution in the case 
of the ATLAS fiducial volume compared to the full phase space case. 
In the case of full phase space (left panel) the four-pion system is created
with relatively large longitudinal momenta. For the ATLAS cuts (right panel) 
the four-pion system is almost at rest and the whole available energy
is transferred to the excitation of the four-pion system.

\subsection{Interference effect}
\begin{figure}[htp]
  \centering
  \includegraphics[width=0.5\textwidth, angle = -90]{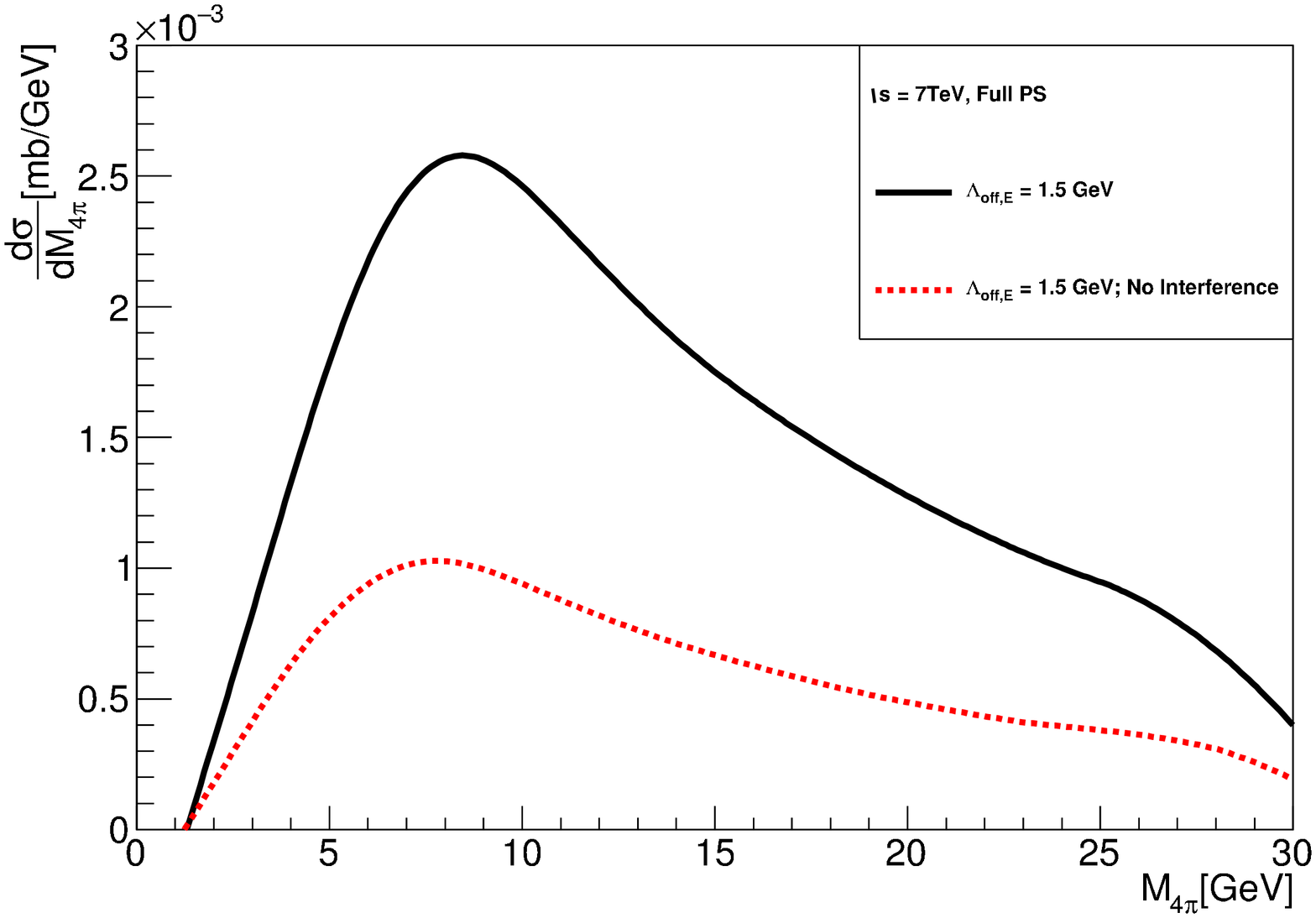}
  \includegraphics[width=0.5\textwidth, angle = -90]{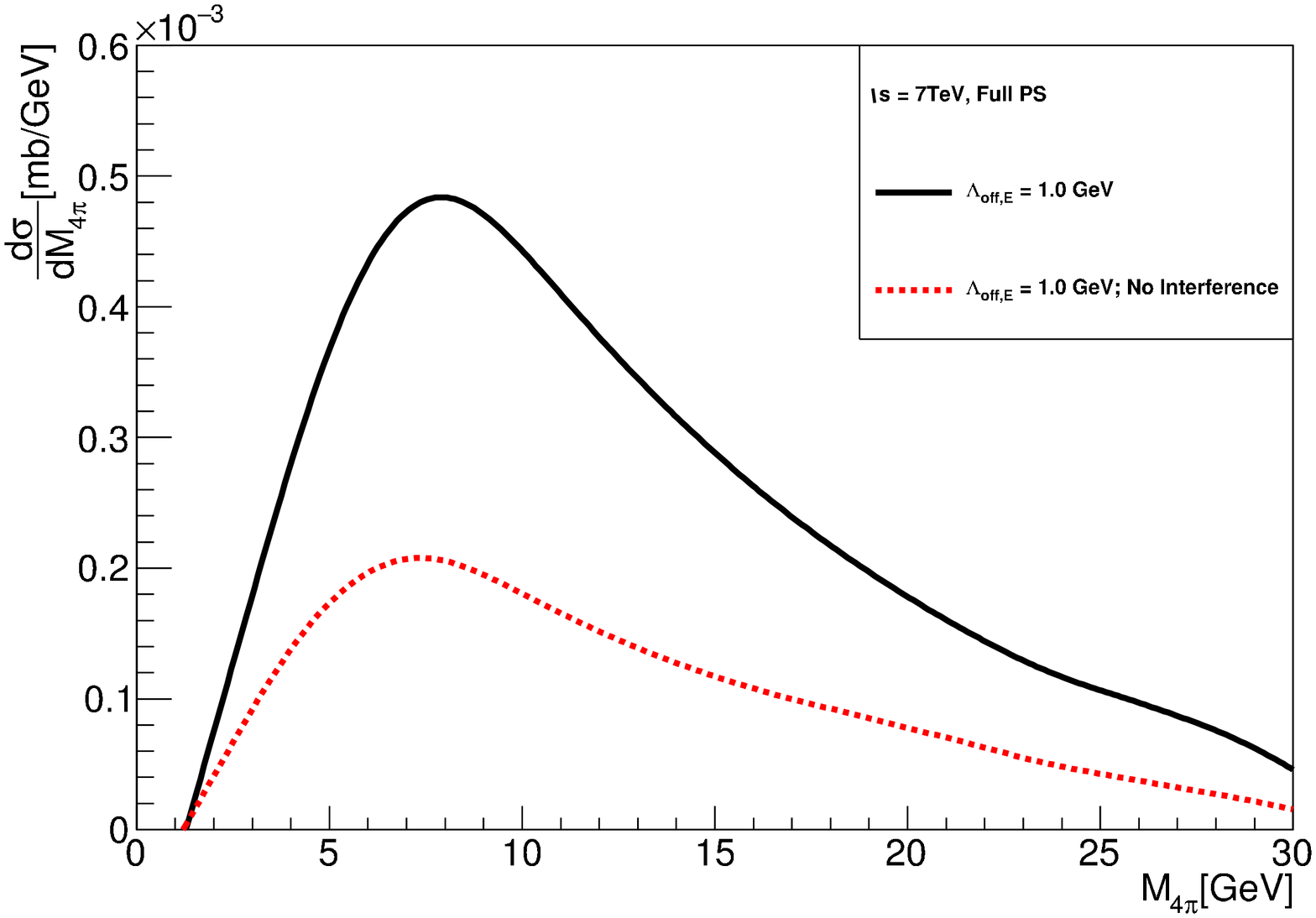}
  \caption{Four-pion invariant mass distribution for the full phase
    space for two different values of $\Lambda_{off,E}$.
}
  \label{PH_CM_interference1Lambda}
\end{figure}

In this subsection we investigate interference between graphs in the whole amplitude. In order to quantify this effect
we propose to compare the cross section for the full amplitude of (\ref{sum_of_diagrams}), with the cross section obtained by adding matrix element squared of individual diagrams, i.e., 
\begin{equation}
\begin{split}
|{\cal M}_{\textrm{no interference}}|^{2} &= \frac{1}{4}\left(|{\cal M}_{\{3456\}}|^{2} + |{\cal M}_{\{5436\}}|^{2} + |{\cal M}_{\{3654\}}|^{2} 
+ |{\cal M}_{\{5634\}}|^{2}
\right)+\ldots.\\
\label{sum_of_diagrams_noInterference} 
\end{split}
\end{equation}
The plots for the full phase space are presented in Fig. \ref{PH_CM_interference1Lambda}.
The amount of the interference can be considered as a measure of rapidity ordering characteristic for high energy multiperipheral processes. For fully ordered events i.e. with large rapidity gaps between all particles the interference effect is small because identical particles occupy different regions of phase space and the amplitude with reversed order is damped by the factor responsible for peripherality of the process. In our case the identical pions are often spaced by the large  rapidity gap (see Figs.\ref{Full_MCentral} and \ref{Atlas_MCentral}), however, low rapidity gap spacing component is also strong. As result interference effect contributes to $\sim 1/2$ of the cross section as seen in Fig. \ref{PH_CM_interference1Lambda}. 

\subsection{Rapidity ordering of pions and the gap between two pion systems}
\begin{figure}[htp]
  \centering
  \includegraphics[width=0.40\textwidth, angle = -90]{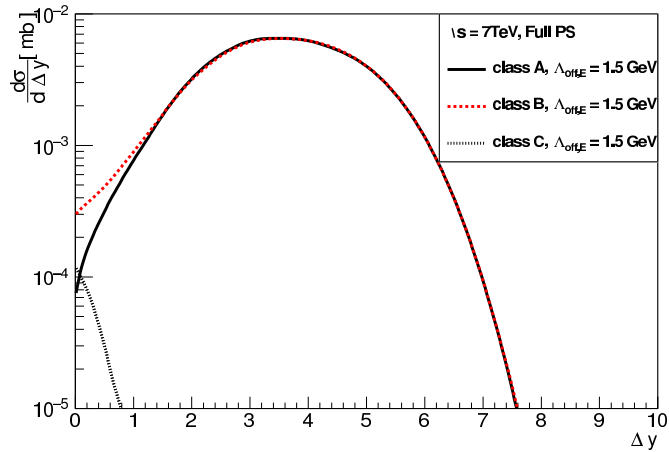}
  \caption{Rapidity difference between second and the third pion for $\Lambda_{off,E}=1.5$~GeV for full phase space.}
\label{Full_MCentral}
\end{figure}

\begin{figure}[htp]
  \centering
  \includegraphics[width=0.40\textwidth, angle = -90]{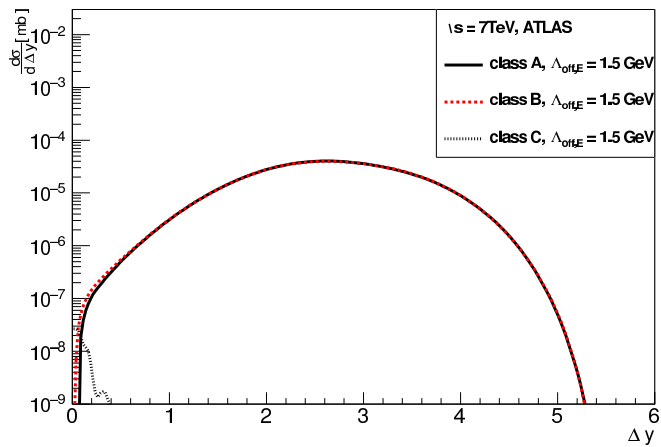}
  \caption{Rapidity difference between second and the third pion for $\Lambda_{off,E}=1.5$~GeV for the ATLAS detector cuts.}
\label{Atlas_MCentral}
\end{figure}

In this subsection the rapidity gap between different orderings in rapidity of pions will be presented. This variable can well distinguish different central particles, however, any other variable that separates pions can be used. The procedure can be used in experiment to characterize triple pomeron/reggeon exchange process. The idea is as follows. The pions will be ordered with respect to their rapidities. Assume that the rapidities of pions are ordered in the following way $y_{1} < y_{2} < y_{3} < y_{4}$. The distribution in rapidity difference between pions $2$ and $3$ will be presented. Three different classes of the ordering of pion charges are possible in general. The class $A$:
\begin{itemize}
 \item {$\pi^{+}(y_{1})$, $\pi^{-}(y_{2})$, $\pi^{+}(y_{3})$, $\pi^{-}(y_{4})$,}
  \item {$\pi^{-}(y_{1})$, $\pi^{+}(y_{2})$, $\pi^{-}(y_{3})$, $\pi^{+}(y_{4})$,}
\end{itemize}
the class $B$:
 \begin{itemize}
  \item {$\pi^{-}(y_{1})$, $\pi^{+}(y_{2})$, $\pi^{+}(y_{3})$, $\pi^{-}(y_{4})$,}
  \item {$\pi^{+}(y_{1})$, $\pi^{-}(y_{2})$, $\pi^{-}(y_{3})$, $\pi^{+}(y_{4})$,}
 \end{itemize}
 and the class $C$:
 \begin{itemize}
 \item {$\pi^{+}(y_{1})$, $\pi^{+}(y_{2})$, $\pi^{-}(y_{3})$, $\pi^{-}(y_{4})$,}
 \item {$\pi^{-}(y_{1})$, $\pi^{-}(y_{2})$, $\pi^{+}(y_{3})$, $\pi^{+}(y_{4})$.}
\end{itemize}
In Figs. \ref{Full_MCentral} and \ref{Atlas_MCentral} we present distributions in rapidity difference between the second and the third pion for full phase space and for the ATLAS kinematical cuts (\ref{ATLASCut1}). These plots show the characteristics of the triple-Regge process which can be verified experimentally.
As can be seen from the figures, the events for the class $C$ happens much more rarely than the events for classes $A$ or $B$. In addition, the gap for the class $C$ is much smaller than for classes $A$ and $B$. This is because pion exchange is responsible for the gap for the class $C$ versus pomeron exchange for classes $A$ and $B$.


\subsection{Comparison with $2\sigma$ production}
The $pp \rightarrow pp \sigma\sigma$ process recently discussed in \cite{LNS2016_4pi}, due to the decay $\sigma \rightarrow \pi^{+} \pi^{-}$, produces the same final state as the triple-Regge $pp \rightarrow pp4\pi$ process.

The Born-level results for the continuum mechanism including ATLAS cuts
(\ref{ATLASCut1}) 
for $\sqrt{s}=7$ and 13~TeV, see Tab. \ref{tab:CrossSection13TeV}, should be compared to 
750.56~nb and 847.46~nb, respectively, from the sequential $pp \to pp (\sigma
\sigma \to 4 \pi)$ 
process discussed in \cite{LNS2016_4pi} (please note that a slightly different cuts were used there).
Note that these values of cross section are smaller than in 
Table~I of \cite{LNS2016_4pi} where $p_{t,\pi} > 0.1$~GeV was imposed in the
calculations.
The cross sections for the $\sigma \sigma$ mechanism was obtained with
the coupling constants given by (2.12) of \cite{LNS2016_4pi}
and the off-shell $t$-channel $\sigma$ meson form factor (2.13) of
\cite{LNS2016_4pi} 
with $\Lambda_{off,E} = 1.6$~GeV.

\begin{figure}[htp]
  \centering
  \includegraphics[width=0.5\textwidth, angle = -90]{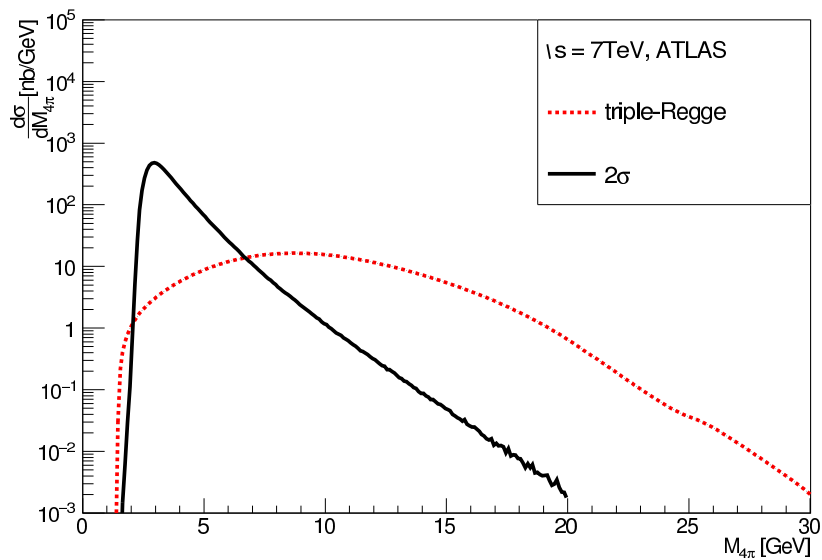}
  \caption{Four-pion invariant mass distribution ($M_{4 \pi}$) 
          with the ATLAS kinematical cuts (\ref{ATLASCut1}) for $\sqrt{s}=7$ TeV. The results correspond to the Born level calculations. The dotted line represents the triple Regge exchange mechanism obtained for $\Lambda_{off,E}=1.5$~GeV. The solid line represents the contribution from $\sigma\sigma$ mechanism discussed in \cite{LNS2016_4pi}.}
\label{ATLAS_CMRhoComaprison}
\end{figure}

These two mechanisms are complementary as can be seen in Fig. \ref{ATLAS_CMRhoComaprison}, as they occupy different range of $M_{4\pi}$. In addition, in the range $2{\rm ~GeV} < M_{4\pi} < 4 {\rm ~GeV}$ the $\sigma\sigma$ mechanism dominates with a rather sharp peak at $M_{4\pi} \sim 3$~GeV and the triple-Regge contribution dominates above $8$ GeV. These characteristics could be very useful when trying experimental distinctions of these two processes.

\subsection{Predictions for LHC at $13$ TeV}
\label{Subsection:13TeV}
In this subsection we wish to provide also first predictions for current runs at the LHC at $\sqrt{s}=13$~TeV. A more detailed analysis, including technical details of experiments, will be postponed to a separate paper.

In Tab. \ref{tab:CrossSection13TeV} numerical values of the cross section are given and compared to out previous results.
\begin{table}[!htb]
  \caption{The integrated Born level (no absorption effects) cross section for the four-pion continuum production. Results were calculated for two different values of the cut-off parameter $\Lambda_{off,E}$ (\ref{off-shell_formfactor}).}
\centering
  \begin{tabular}{| l | c | c | c | }
    \hline
    & $\Lambda_{off,E}$ [GeV] & $\sigma$ $@$ $\sqrt{s}=7$ TeV & $\sigma$ $@$ $\sqrt{s}=13$ TeV  \\ \hline    
    Full PS    & 1.0 & 7.21  $\mu$b & 8.97  $\mu$b \\ \hline
    Full PS   & 1.5 & 42.86 $\mu$b & 51.78 $\mu$b \\ \hline
    ATLAS   & 1.0 & 6.91   nb     &  7.48 nb \\ \hline
    ATLAS   & 1.5 & 141.43 nb     &  154.19 nb \\ \hline
    ALICE   & 1.0 & 4.2  pb     & 4.7 pb  \\ \hline
    ALICE   & 1.5 & 37.7 pb     & 42 pb  \\ \hline
  \end{tabular}
  \label{tab:CrossSection13TeV}
\end{table}
The table shows that the transfer of energy to the system is slowly varying with the collision energy. Therefore all plots presented in the previous sections do not differ dramatically for the case of $\sqrt{s}=13$~TeV. The only sizeable difference is that in the rapidity plots the protons are a bit further from $y=0$. Summing up, the model cross section is only weakly dependent on the centre-of-mass energy.

\section{Conclusions}

The triple-Regge exchange model  was proposed for 
the production of four-pions in the $p p \to p p \pi^+ \pi^- \pi^+ \pi^-$ exclusive reaction. The amplitudes of the process were parametrized
in the Regge formalism with coupling constants fixed to describe 
the total nucleon-nucleon and pion-nucleon cross sections.
Some care must be taken how to 'remove' the low dipion invariant mass regions
($M_{\pi\pi} < 2$~GeV) that are not described by Regge amplitudes

In the considered process two of the pions
are off-mass-shell already in the Born amplitude(s). The off-shell effects
are parametrized in terms of corresponding form factors.
The same objects (form factors) were discussed recently 
in the context of the $p p \to p p \pi^+ \pi^-$ reaction considered 
both theoretically as well as measured by the STAR, CDF, and CMS collaborations \cite{STAR, CDF, CMS}. 
The cut-off parameter was fitted then \cite{LNS2016_2pi} 
to describe the preliminary data. The present dipion data do not allow
for a precise extraction of the model parameter but allow to
obtain a reasonable range of the cut-off parameter 
$\Lambda_{off,E}$ = 1 -- 1.5 GeV.
Here we have assumed exponential dependence of the form factors 
on the pion virtualities.
Then the model has almost only one free parameter 
(called here cut-off parameter), which can be taken 
in the range known from the four-body ($p p \to p p \pi^+ \pi^-$) reaction 
studied in the literature.
In comparison to the four-body reaction the dependence on the cut-off
parameter is much stronger as two pions, instead of one for the $pp \rightarrow pp\pi^{+}\pi^{-}$ process, are off-mass-shell.

We have made first predictions for the six-body processes.
Both total cross sections (integrated over six-body phase space)
as well as several differential distributions were calculated and
presented. Compared to the $\sigma \sigma$ and $\rho \rho$
mechanisms considered recently by two of us \cite{LNS2016_4pi}, 
the considered here mechanism populates final states with much larger  
dipion and four-pion invariant masses. 
We get total cross section 7.21 -- 42.86~$\mu$b (see Tab. \ref{tab:CrossSection13TeV})
in the whole phases space (neglecting absorption effects!).
The absorption effects are expected to diminish the cross section by an order of magnitude. 
Our preliminary studies here have been done at the Born level and the
absorption can be included only in the form of the multiplicative 
gap survival factor. One expects it to be of the order of 0.1. Full-fledged calculation of absorption effects and in particular 
its dependence on kinematical variables is not simple 
(see, e.g., \cite{LS2015} for detailed studies for the $p p \to p p \pi^+ \pi^-$ reaction).

The integrated full phase space cross section cannot be, however, 
measured due to limited coverage of the LHC detectors.
We have therefore made predictions for the kinematical cuts
characteristic for the ATLAS and ALICE detectors.
The latter detector can identify pions down to very small transverse momenta
of $p_{t,\pi}$ = 0.1 GeV. However, the rapidity coverage of the ALICE
tracker is very (too) limited. This does not allow to observe the large
four-pion invariant masses, the genuine feature of the considered diffractive triple-Regge mechanism. 
In contrast, the ATLAS detector allows to measure cases
with large $4 \pi$ invariant masses. We expect that the considered
multi-diffractive process dominates over the contributions of other 
mechanisms for four-pion invariant masses $M_{4 \pi} > 10$~GeV. 

We have discussed in addition how much energy can be transferred from protons to 
the excitation of the four-pion system. We have demonstrated that the model amplitude gives natural limitations for such a transfer.
A specific ordering of pion charges in rapidity has been found to be an interesting and representative characteristics of the discussed process.

To assure exclusivity of the process, not only charged pions 
but also forward/backward protons should be measured. 
The ALFA detectors are natural candidates 
for this purpose in the case of the ATLAS experiments. 
Similarly the CMS collaboration together with the TOTEM collaboration 
could perform similar studies.

In summary, the observation of counts/events at large four-pion invariant masses
should be a clear signal of observing the discussed here three-pomeron exchange processes, 
not identified so far experimentally.

\acknowledgments
This work was supported in part by the Polish National Science Centre Grant
No. 2014/15/B/ST2/02528, the Ministry of Science and Higher Education Republic of Poland Grant No. IP2014~025173 (Iuventus Plus)
and by the Center for Innovation and Transfer of Natural Sciences 
and Engineering Knowledge in Rzesz{\'o}w. This research was supported in part by PLGrid Infrastructure. Some calculations were also 
supported by Cracow Cloud One infrastructure. 

\appendix

\section{Peripheral reaction with decay of central system}
\label{A}

In this section (appendix) we presents the recipe for generating
phase space of the reaction which treats all final pions in the same way
and therefore it is suitable for the case when including
complicated interferences of contributing amplitudes. 
The considered reaction is of the form: 
$p(p_{a})+p(p_{b}) \rightarrow p(p_{1})+p(p_{2}) + CM(P_{4\pi})$ and then 
$CM(P_{4\pi})\rightarrow \pi^{+}(p_{3})+\pi^{-}(p_{4}) + \pi^{+}(p_{5})+\pi^{-}(p_{6})$.
The formula for the cross section can be written in the standard form
\begin{equation}
\sigma = \int (2\pi)^{4} \frac{\overline{|{\cal M}|^{2}}}{2s}
\delta^{(4)}\left(P-\sum_{i=1}^{6}p_{i}\right)
\prod_{i=1}^{6} \frac{d^{3}p_{i}}{(2\pi)^{3}2E_{i}}\,,
 \label{sigmaGeneral}
\end{equation}
where ${\cal M}$ is a matrix element for the six-body reaction
and $P = p_{a}+p_{b}$ is a total four-momentum in the initial (and final) system.

Starting from (\ref{sigmaGeneral}) the phase space 
is factorized as (see \cite{Pilkuhn}, Eq. (9.7))
\begin{equation}
\begin{split}
&\sigma = \int_{4m_{\pi}}^{\infty} dM_{4\pi}^{2} \int (2\pi)^{4} \delta^{(4)}\left(P-p_{1}-p_{2}-P_{4\pi}\right) \frac{d^{3}p_{1}}{(2\pi)^{3}2E_{1}} \frac{d^{3}p_{2}}{(2\pi)^{3}2E_{2}} \frac{d^{3}P_{4\pi}}{(2\pi)^{3}2E_{4\pi}} \\
& \qquad \times \frac{1}{2\pi}\int (2\pi)^{4} \delta^{(4)}(P_{4\pi} - \sum_{i=3}^{6}
p_{i})  \prod_{i=3}^{6} \frac{d^{3}p_{i}}{(2\pi)^{3}2E_{i}}
\frac{\overline{|{\cal M}|^{2}}}{2s}
\,,
\end{split}
\label{sigma_in_appendix}
\end{equation}
where the integration over $M_{4\pi}$ variable extends from the threshold for the $4\pi$ production to the infinity (in our case to the technical cut $M_{4\pi}<30$ GeV).
The second decay of the central system $P_{4\pi}$ into four particles can be calculated using
slightly modified sequence of decays of the GENBOD CERN library 
(currently the TGenPhaseSpace class from ROOT package \cite{ROOT_site}).
For a description of the algorithm of the generation see \cite{JamesCERN}. 
This modification will be described elsewhere.

This prescription is the best choice for matrix elements with
permutation of identical particles, 
as it treats all centrally produced particles on the same footing.

In our practical realization the phase space available for the process is fairy large which requires special technical treatment event for adaptive Monte Carlo generator. 
The most efficient solution is to divide the whole range of $M_{4\pi}$ into smaller exclusive intervals and add distributions for the different intervals.


\end{document}